\documentclass[twocolumn, trackchanges]{aastex62}
\submitjournal{ApJ}
\usepackage{comment}

\shorttitle{Strong  (H$\beta$+[\ion{O}{3}]) and H$\alpha$ emitters at redshift $z \simeq 7 - 8$.}
\shortauthors{Rinaldi et al.}

\begin{document}

\title{MIDIS: Strong  (H$\beta$+[\ion{O}{3}]) and H$\alpha$ emitters at redshift $z \simeq 7 - 8$ unveiled with JWST/NIRCam and MIRI imaging in the Hubble eXtreme Deep Field (XDF)}

\newcommand{\gsim}{{\;\raise0.3ex\hbox{$>$\kern-0.75em\raise-1.1ex\hbox{$\sim$}}\;}}

\correspondingauthor{Pierluigi Rinaldi}
\email{rinaldi@astro.rug.nl}

\author[0000-0002-5104-8245]{P. Rinaldi}
\affiliation{Kapteyn Astronomical Institute, University of Groningen,
P.O. Box 800, 9700AV Groningen,
The Netherlands}

\author[0000-0001-8183-1460]{K. I. Caputi}
\affiliation{Kapteyn Astronomical Institute, University of Groningen,
P.O. Box 800, 9700AV Groningen,
The Netherlands}

\affiliation{Cosmic Dawn Center (DAWN), Denmark}

\author[0000-0001-6820-0015]{L. Costantin}
\affiliation{Centro de Astrobiolog\'{\i}a (CAB), CSIC-INTA, Ctra. de Ajalvir km 4, Torrej\'on de Ardoz, E-28850, Madrid, Spain}

\author[0000-0001-9885-4589]{S. Gillman}
\affiliation{Cosmic Dawn Center (DAWN), Denmark}

\affiliation{DTU-Space, Elektrovej, Building 328 , 2800, Kgs. Lyngby, Denmark}

\author[0000-0001-8386-3546]{E. Iani}
\affiliation{Kapteyn Astronomical Institute, University of Groningen, P.O. Box 800, 9700AV Groningen, The Netherlands}

\author[0000-0003-4528-5639]{P. G. P\'erez-Gonz\'alez}
\affiliation{Centro de Astrobiolog\'{\i}a (CAB), CSIC-INTA, Ctra. de Ajalvir km 4, Torrej\'on de Ardoz, E-28850, Madrid, Spain}

\author[0000-0002-3005-1349]{G. \"Ostlin}
\affiliation{Department of Astronomy, Stockholm University, Oscar Klein Centre, AlbaNova University Centre, 106 91 Stockholm, Sweden}

\author[0000-0002-9090-4227]{L. Colina}
\affiliation{Centro de Astrobiolog\'{\i}a (CAB), CSIC-INTA, Ctra. de Ajalvir km 4, Torrej\'on de Ardoz, E-28850, Madrid, Spain}
\affiliation{Cosmic Dawn Center (DAWN), Denmark}

\author[0000-0002-2554-1837]{T. R. Greve}
\affiliation{Cosmic Dawn Center (DAWN), Denmark}
\affiliation{DTU-Space, Elektrovej, Building 328 , 2800, Kgs. Lyngby, Denmark}

\author[0000-0000-0000-0000]{H. U. Noorgard-Nielsen}
\affiliation{Cosmic Dawn Center (DAWN), Denmark}
\affiliation{DTU-Space, Elektrovej, Building 328 , 2800, Kgs. Lyngby, Denmark}

\author[0000-0000-0000-0000]{G. S. Wright}
\affiliation{UK Astronomy Technology Centre, Royal Observatory Edinburgh,
Blackford Hill, Edinburgh EH9 3HJ, UK}

\author[0000-0001-6794-2519]{A. Alonso-Herrero}
\affiliation{Centro de Astrobiolog\'{\i}a (CAB), CSIC-INTA, Camino Bajo del Castillo s/n, E-28692 Villanueva de la Ca\~nada, Madrid, Spain}

\author[0000-0002-7093-1877]{J. \'Alvarez-M\'arquez}
\affiliation{Centro de Astrobiolog\'{\i}a (CAB), CSIC-INTA, Ctra. de Ajalvir km 4, Torrej\'on de Ardoz, E-28850, Madrid, Spain}

\author[0000-0000-0000-0000]{A. Eckart}
\affiliation{I.Physikalisches Institut der Universit\"at zu K\"oln, Z\"ulpicher Str. 77,
50937 K\"oln, Germany}

\author[0000-0000-0000-0000]{M. Garc\'{\i}a-Mar\'{\i}n}
\affiliation{European Space Agency/Space Telescope Science Institute, 3700 San Martin Drive, Baltimore MD 21218, USA}

\author[0000-0002-4571-2306]{J. Hjorth}
\affiliation{DARK, Niels Bohr Institute, University of Copenhagen, Jagtvej 128,
2200 Copenhagen, Denmark}

\author[0000-0002-7303-4397]{O. Ilbert}
\affiliation{Aix Marseille Universit\'e, CNRS, LAM (Laboratoire d’Astrophysique de Marseille) UMR 7326, 13388, Marseille, France}

\author[0000-0002-7612-0469]{S. Kendrew}
\affiliation{European Space Agency/Space Telescope Science Institute, 3700 San Martin Drive, Baltimore MD 21218, USA}

\author[0000-0002-0690-8824]{A. Labiano}
\affiliation{Telespazio UK for the European Space Agency (ESA), ESAC, Camino Bajo del Castillo s/n, 28692 Villanueva de la Ca\~nada, Spain}

\author[0000-0000-0000-0000]{O. Le F\`evre}
\affiliation{Aix Marseille Universit\'e, CNRS, LAM (Laboratoire d’Astrophysique de Marseille) UMR 7326, 13388, Marseille, France}

\author[0000-0002-0932-4330]{J. Pye}
\affiliation{School of Physics \& Astronomy, Space Research Centre, Space Park Leicester, University of Leicester, 92 Corporation Road, Leicester, LE4 5SP, UK}

\author[0000-0000-0000-0000]{T. Tikkanen}
\affiliation{School of Physics \& Astronomy, Space Research Centre, Space Park Leicester, University of Leicester, 92 Corporation Road, Leicester, LE4 5SP, UK}

\author[0000-0003-4793-7880]{F. Walter}
\affiliation{Max-Planck-Institut f\"ur Astronomie, K\"onigstuhl 17, 69117 Heidelberg, Germany}

\author[00000-0001-5434-5942]{P. van der Werf}
\affiliation{Leiden Observatory, Leiden University, PO Box 9513, 2300 RA Leiden, The Netherlands}

\author[0000-0003-1810-0889]{M. Ward}
\affiliation{Centre for Extragalactic Astronomy, Durham University, South
Road, Durham DH1 3LE, UK}

\author[0000-0002-8053-8040]{M. Annunziatella}
\affiliation{Centro de Astrobiolog\'{\i}a (CAB), CSIC-INTA, Ctra. de Ajalvir km 4, Torrej\'on de Ardoz, E-28850, Madrid, Spain}
\affiliation{INAF-Osservatorio Astronomico di Capodimonte, Via Moiariello 16, I-80131 Napoli, Italy}

\author[0000-0002-0438-0886]{R. Azzollini}
\affiliation{Centro de Astrobiolog\'{\i}a (CAB), CSIC-INTA, Ctra. de Ajalvir km 4, Torrej\'on de Ardoz, E-28850, Madrid, Spain}
\affiliation{Dublin Institute for Advanced Studies, Astronomy \& Astrophysics Section, 31 Fitzwilliam Place, Dublin 2, Ireland}

\author[0000-0001-8068-0891]{A. Bik}
\affiliation{Department of Astronomy, Stockholm University, Oscar Klein Centre, AlbaNova University Centre, 106 91 Stockholm, Sweden}

\author[0000-0002-3952-8588]{L. Boogaard}
\affiliation{Max-Planck-Institut f\"ur Astronomie, K\"onigstuhl 17, 69117 Heidelberg, Germany}

\author[0000-0001-8582-7012]{S. E. I. Bosman}
\affiliation{Max-Planck-Institut f\"ur Astronomie, K\"onigstuhl 17, 69117 Heidelberg, Germany}

\author[0000-0003-2119-277X]{A. Crespo G\'{o}mez}
\affiliation{Centro de Astrobiolog\'{\i}a (CAB), CSIC-INTA, Ctra. de Ajalvir km 4, Torrej\'on de Ardoz, E-28850, Madrid, Spain}

\author[0000-0002-2624-1641]{I. Jermann}
\affiliation{Cosmic Dawn Center (DAWN), Denmark}
\affiliation{DTU-Space, Elektrovej, Building 328 , 2800, Kgs. Lyngby, Denmark}

\author[0000-0001-5710-8395]{D. Langeroodi}
\affiliation{DARK, Niels Bohr Institute, University of Copenhagen, Jagtvej 128, 2200 Copenhagen, Denmark}

\author[0000-0003-0470-8754]{J. Melinder}
\affiliation{Department of Astronomy, Stockholm University, Oscar Klein Centre, AlbaNova University Centre, 106 91 Stockholm, Sweden}

\author[0000-0001-5492-4522]{R. A. Meyer}
\affiliation{Max-Planck-Institut f\"ur Astronomie, K\"onigstuhl 17, 69117 Heidelberg, Germany}

\author[0000-0002-3305-9901]{T. Moutard}
\affiliation{Aix Marseille Universit\'e, CNRS, LAM (Laboratoire d’Astrophysique de Marseille) UMR 7326, 13388, Marseille,
France}

\author[0000-0000-0000-0000]{F. Peissker}
\affiliation{I.Physikalisches Institut der Universit\"at zu K\"oln, Z\"ulpicher Str. 77,
50937 K\"oln, Germany}

\author[0000-0000-0000-0000]{M. Topinka}
\affiliation{Dublin Institute for Advanced Studies, Astronomy \& Astrophysics Section, 31 Fitzwilliam Place, Dublin 2, Ireland}

\author[0000-0001-7591-1907]{E. van Dishoeck}
\affiliation{Leiden Observatory, Leiden University, PO Box 9513, 2300 RA Leiden, The Netherlands}

\author[0000-0001-9818-0588]{M. G\"udel}
\affiliation{Dept. of Astrophysics, University of Vienna, Türkenschanzstr 17, A-1180 Vienna, Austria}
\affiliation{Max-Planck-Institut für Astronomie (MPIA), Königstuhl 17, 69117 Heidelberg, Germany}
\affiliation{ETH Zürich, Institute for Particle Physics and Astrophysics, Wolfgang-Pauli-Str. 27, 8093 Zürich, Switzerland}

\author[0000-0002-1493-300X]{Th. Henning}
\affiliation{Max-Planck-Institut f\"ur Astronomie, K\"onigstuhl 17, 69117 Heidelberg, Germany}

\author[0000-0000-0000-0000]{P.-O. Lagage}
\affiliation{AIM, CEA, CNRS, Universit\'e Paris-Saclay, Universit\'e Paris
Diderot, Sorbonne Paris Cit\'e, F-91191 Gif-sur-Yvette, France}

\author[0000-0000-0000-0000]{T. Ray}
\affiliation{Dublin Institute for Advanced Studies, Astronomy \& Astrophysics Section, 31 Fitzwilliam Place, Dublin 2, Ireland}

\author[0000-0000-0000-0000]{B. Vandenbussche}
\affiliation{Institute of Astronomy, KU Leuven, Celestijnenlaan 200D bus 2401,
3001 Leuven, Belgium}

\author[0000-0000-0000-0000]{C. Waelkens}
\affiliation{Institute of Astronomy, KU Leuven, Celestijnenlaan 200D bus 2401,
3001 Leuven, Belgium}

\author[0000-0001-6066-4624]{R. Navarro-Carrera}
\affiliation{Kapteyn Astronomical Institute, University of Groningen, P.O. Box 800, 9700AV Groningen, The Netherlands}

\author[0000-0002-5588-9156]{V. Kokorev}
\affiliation{Kapteyn Astronomical Institute, University of Groningen, P.O. Box 800, 9700AV Groningen, The Netherlands}

\begin{abstract}
  We make use of \textit{JWST} medium and broad-band NIRCam imaging, along with ultra-deep MIRI $5.6 \rm \mu m$ imaging, in the Hubble eXtreme Deep Field (XDF) to identify prominent line emitters at $z\simeq 7-8$. Out of a total of 58 galaxies at $z\simeq 7-8$, we find 18 robust candidates ($\simeq$31\%) for (H$\beta$ + [\ion{O}{3}]) emitters, based on their enhanced fluxes in the F430M and F444W filters, with EW$_{0}$(H$\beta$ +[\ion{O}{3}]) $\simeq 87 - 2100 \, \rm \AA$.  Among these emitters, 16 lie in the MIRI coverage area and 12 exhibit a clear flux excess at $5.6 \, \rm \mu m$, indicating the simultaneous presence of a prominent H$\alpha$ emission line with EW$_{0}$(H$\alpha$) $\simeq 200-3000 \, \rm \AA$. This is the first time that  H$\alpha$ emission can be detected in individual galaxies at $z>7$. The H$\alpha$ line, when present,  allows us to separate the contributions of H$\beta$  and [\ion{O}{3}] to the (H$\beta$ +[\ion{O}{3}]) complex, and derive H$\alpha$-based star formation rates (SFRs). We find that in most cases [\ion{O}{3}]/H$\beta > 1$.  Instead, two galaxies have [\ion{O}{3}]/H$\beta < 1$, indicating that the NIRCam flux excess is mainly driven by H$\beta$. This could potentially imply extremely low metallicities. The most prominent line emitters are very young starbursts or galaxies on their way to/from the starburst cloud. They make for a cosmic SFR density $\rm log_{10}(\rho_{SFR_{H\alpha}}/ (\rm M_\odot \, yr^{-1} \, Mpc^{-3})) \simeq -2.35$, which is about a quarter of the total value ($\rm log_{10}(\rho_{SFR_{tot}}/ (\rm M_\odot \, yr^{-1} \, Mpc^{-3})) \simeq -1.76$) at $z\simeq 7-8$. Therefore, the strong H$\alpha$ emitters likely had a significant role in reionization.

\end{abstract}

.
\keywords{Galaxies: formation, evolution,  high-redshift, star formation, starburst, Epoch of Reionization}

\section{Introduction}
Quantifying the presence and properties of galaxies present at the Epoch of Reionization (EoR) is necessary to explain how this major phase transition of the Universe has occurred. Over the past decade, many studies have focused on this topic, but a few important problems complicated the selection of galaxies at this cosmic time. The increasing intergalactic medium absorption with redshift means that basically all photons blueward of the Lyman-$\alpha$ spectral line at $\lambda_{\rm rest} = 1216 \, \rm \AA$ cannot reach us. Indeed, it is well known that the incidence of Lyman-$\alpha$ emitters (LAEs) has a sharp drop at $z>7$ \citep[e.g.,][]{fontana+10,ono+12,caruana+14,pentericci+14}.  Therefore, other emission lines at longer wavelengths must be considered to facilitate the search of galaxies at such high redshifts \citep[e.g.,][]{stark+15}.

However, detecting the optical emission from atomic transitions at $z>7$ was virtually impossible until now, given the lack of sufficiently sensitive near and mid-infrared observatories. The recent advent of the \textit{JWST} is now radically changing this situation by offering, for the first time, sensitive imaging and spectroscopy at such long wavelengths. Indeed, in the first six months of operations, \textit{JWST} has enabled a number of studies of $z>7$ galaxies, particularly on their line emission properties \citep[e.g.,][]{Arellano_2022, Langeroodi_2022, Morishita_2022,Trump_2022, Wang_2022, Williams_2022}.

With imaging, the search of line emitters is facilitated by the fact that the rest-frame equivalent widths (EWs$_{0}$) of some of the main optical emission lines appear to increase, on average, with redshift \citep[e.g., ][]{debarros_2019, Matthee_2023}.  This has allowed for the search of prominent line emitters at intermediate and high redshifts, by identifying galaxies with photometric excess in narrow-band images \citep[e.g.,][]{Khostovan_2016} and even broad-band images \citep[e.g.,][]{Faisst_2016, Roberts-Borsani_2016, Smit_2016, Caputi_2017}. This trend of increasing EWs$_{0}$ with redshift is indicative of an evolution in the galaxy average specific star formation rates (sSFR)  \citep[e.g., ][]{Faisst_2016, Tang_2019}, as well as the conditions of their interstellar medium (ISM) \citep[e.g., ][]{Schaerer_2009}.

At $z>7$  both the H$\beta \, \lambda4861 \, \rm \AA$  and [\ion{O}{3}]~$\lambda\lambda 4959, 5007$ emission lines are shifted into the \textit{JWST's} Near-IR Camera \citep[NIRCam; ][]{Rieke_2005}  wavelength range, making that these lines together can produce a flux excess in the NIRCam filters at $\simeq 4-5 \, \rm \mu m$.  In turn, the (H$\alpha \, \lambda 6563$ + [NII]~$\lambda\lambda  6548, 6583$ + [SII]~$\lambda\lambda 6716,6730$) complex appears in the Mid-Infrared Instrument \citep[MIRI; ][]{Rieke_2015, Wright_2015} wavelength domain at observed $> 5 \, \rm \mu m$. 

In this paper, we make use of publicly available NIRCam images in the Hubble eXtreme Deep Field (XDF) to search for (H$\beta$ + [\ion{O}{3}]) emitters at $z \simeq 7-8$. In most of this field, we also benefit from ultra-deep MIRI $5.6 \, \rm \mu m$ imaging, which we analyse to search for the presence of H$\alpha$  emission in the same galaxies. This is the first time that the H$\alpha$ line can be detected and quantified in individual galaxies at $z>7$. This paper is organised as follows: in \S\ref{sec_2} we describe the datasets, photometric measurements and spectral energy distribution (SED) fitting that allows us to select galaxies at $z \simeq 7-8$. In \S\ref{sec_3} we explain our methodology to identify strong (H$\beta$ + [\ion{O}{3}]) and  H$\alpha$ emitters amongst these galaxies. We present all our results in \S\ref{sec_4} and our conclusions in \S\ref{sec_5}. Throughout this paper, we consider a cosmology with $\rm H_0=70 \,{\rm km \, s^{-1} Mpc^{-1}}$, $\rm \Omega_M=0.3$ and $\rm \Omega_\Lambda=0.7$. All magnitudes are total and refer to the AB system \citep{Oke_1983}. A \citet{Chabrier_2003} initial mass function (IMF) is assumed.

\section{Datasets, Photometry and SED fitting}\label{sec_2}
\subsection{Datasets}
\begin{figure*}[ht!]
    \centering
    \includegraphics[width = \textwidth]{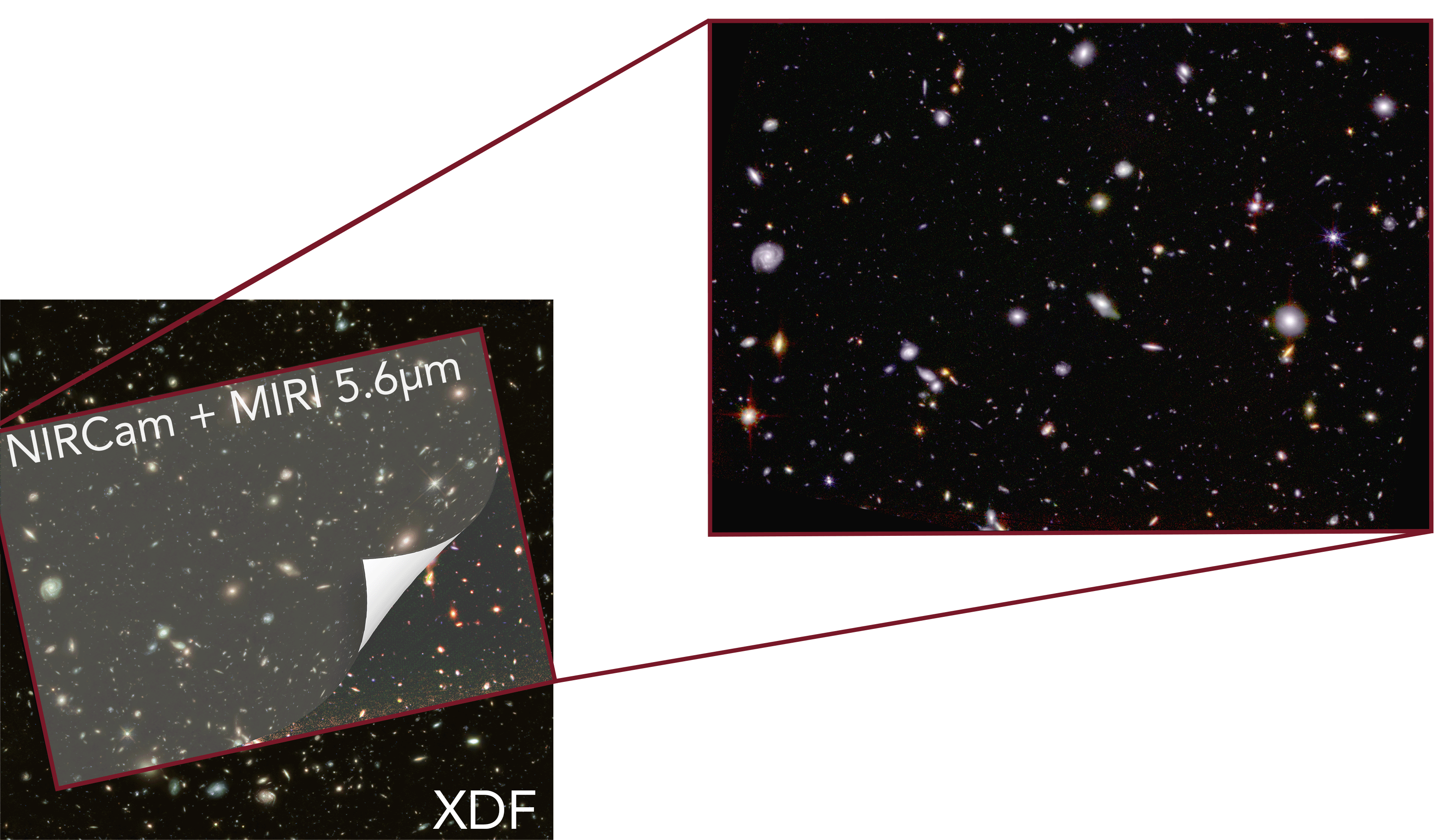}
    \caption{An RGB mosaic of the Hubble eXtreme Ultra Deep Field. This image has been obtained by exploiting the \textit{HST} and \textit{JWST} images currently available in this field. In particular, the background image has been obtained by combining \textit{HST} and \textit{JWST}/NIRCam filters. The zoom-in shows the region covered by MIRI/F560W. In this case, to create the RGB image, we adopted the \textit{JWST} filters only where R: F560W; G: F430M, F444W, F460M, and F480M; B: F182M and F210M. The RGB images have been produced using the software Trilogy \citep[][]{Coe_2015}. } 
    \label{Fig_1}
\end{figure*}

The Hubble XDF \citep[][see Fig. \ref{Fig_1}]{illingworth2013} is a small field of the sky with the deepest \textit{Hubble Space Telescope} (\textit{HST}) observations ever taken since this telescope started operations more than thirty years ago. This field has been the main window to study the early Universe before the \textit{JWST} advent, with numerous works scientifically exploiting its unique possibilities. Now in the \textit{JWST} era, the \textit{HST} data in the XDF and surroundings are being enhanced with deep imaging and spectroscopy obtained with \textit{JWST/}NIRCam and MIRI, extending the wavelength coverage of high spatial-resolution observations to the mid-infrared.

\subsubsection{JWST/NIRCam}
In this work, we made use of the recent \textit{JWST/}NIRCam images collected by \citet{Williams_2023} in a General Observers Cycle-1 program across the HUDF (PID: 1963; PI: Christina C. Williams). Observations have been taken in 5 \textit{JWST}/NIRCam medium bands: F182M, F210M, F430M, F460M, and F480M. In particular, 7.8 hours of the total integration time have been dedicated to F182M, F210M, and F480M. Instead, only 3.8 hours of observations have been collected for F430M and F460M. In order to complement these datasets, we also made use of the imaging data taken as part of \textit{The First Reionization Epoch Spectroscopic COmplete Survey} \citep[][FRESCO]{FRESCO_1895, Oesc_2023} (PID: 1895; PI: Pascal Oesch). On the one hand, this GO program allowed us to add more depth to F182M and F210M, on the other hand it gave us the opportunity to include F444W in our analysis. 

All \textit{JWST}/NIRCam images have been reduced by adopting a modified version of the official \textit{JWST} pipeline\footnote{The pipeline is available at the following \href{https://github.com/spacetelescope/jwst}{link};} (based on \texttt{jwst 1.8.2} and Calibration Reference Data System pipeline mapping (CRDS; pmap) \texttt{1018})). More detailed information about the reference files is available on the official \href{https://jwst-crds.stsci.edu}{STScI/CRDS} website. 

Compared to the official \textit{JWST} pipeline, our version includes different procedures, following some of the ideas presented in \citet{Bagely_2022}, to deal with the unresolved problems that still affect the official software. In our data reduction, we minimized the impact of the so-called “snowballs”, the \textit{1/f} noise, the “wisps”\footnote{More information about these artefacts at the following \href{https://jwst-docs.stsci.edu/jwst-near-infrared-camera/nircam-features-and-caveats/nircam-claws-and-wisps}{link};}, and the residual cosmic rays. After reducing all the \textit{JWST}/NIRCam images from Williams's and FRESCO programs, we drizzled all the NIRCam calibrated files to 0.03\arcsec/pixel, as the final pixel scale we adopted in this work. All the final images have been aligned to the Hubble Legacy Fields (HLF) catalogue\footnote{The HLF catalogue is available at the following \href{https://archive.stsci.edu/hlsps/hlf/}{link};}.

As a sanity check, we compared the photometry for the brightest sources ($<$ 24 mag) in all the NIRCam filters. To do that, we produced two versions of our final images, with and without the extra steps we employed in our modified version of the official pipeline. Then, we extracted the sources by using the software \texttt{Source Extractor} \citep[SExtractor,][]{SExtractor} and compared their photometry. This test demonstrated that our extra steps do not introduce any kind of systematic effect in the photometry.

\subsubsection{JWST/MIRI}
We complemented the {\it JWST}/NIRCam observations with the MIRI $5.6\mu$m imaging from the {\it JWST} Guaranteed Time Observations (GTO) program: {\it MIRI Deep Imaging Survey} (MIDIS) (PID: 1283, PI: G\"oran \"Ostlin).  
The MIRI observations were carried out in December 2022 and targeted with the broad-band filter F560W the HUDF for a total amount of 50 hours ($\approx 41$ hours on-source), covering an area of about 4.7 arcmin$^2$.
By reaching a median depth of 29.15 mag ($5\sigma$, r = 0.15\arcsec), this set of observations represents the deepest imaging available at $5.6\mu$m to date. A complete description of the data collection and reduction, as well as the source statistics on these  $5.6\mu$m images, will be presented by  \citet{Ostlin_2023}. Here we only summarize the basic information of this data processing.

As in the case of the NIRCam imaging, we adopted a modified version of the official {\it JWST} pipeline to reduce the MIRI data.
In fact, the final products that can be obtained by running the {\it JWST} pipeline are still affected by strong patterns (e.g., vertical striping and background gradients) that impact the scientific quality of the images \cite[e.g.,][]{Iani_2022a}.
To overcome these problems, we added to the pipeline some extra steps at the end of stages 2 and 3 that allowed us to significantly mitigate the intensity of the striping, the background inhomogeneities as well as the noise of the output image. 
A comparison between the F560W magnitude of the brightest galaxies ($< 24$ mag) measured in MIRI images obtained with and without the extra steps ensured that our modified version of the pipeline did not introduce any systematic offset.

Finally, we drizzled the final MIRI image to the same pixel scale adopted for the {\it JWST}/NIRCam images and registered its astrometry to the HLF catalogue.

\subsubsection{Ancillary HST data}
We obtained all of \textit{HST} images over the HUDF from the Hubble Legacy Field GOODS-S (HLF-GOODS-S)\footnote{The \textit{HST} images (0.03\arcsec/pixel) have been downloaded from the following \href{https://archive.stsci.edu/prepds/hlf/}{link};}. The HLF-GOODS-S provides 13 \textit{HST} bands covering a wide range of wavelengths (0.2$\mu$m - 1.6$\mu$m), from the UV (WFC3/UVIS F225W, F275W, and F336W filters), optical (ACS/WFC F435W, F606W, F775W, F814W and F850LP filters), to near-infrared (WFC3/IR F098M, F105W, F125W, F140W and F160W filters). See \citet{Whitaker_2019} for more detailed information on these observations.

\subsection{Photometric Analysis}
We used the software \texttt{SExtractor} to detect the sources and measure their photometry in all the 20 filters available from \textit{HST} and \textit{JWST}, covering a wide range of wavelengths (0.2$\mu$m $-$ 5.6$\mu$m). We used \texttt{SExtractor} in dual-image mode adopting a super-detection image that we created by combining photometric information from different bands. In order to maximize the number of the detected sources, we opted to use a \textit{hot-mode} extraction, as presented in \citet{Galametz_2013} which is well suited to find very faint sources. 

We combined aperture photometry, adopting circular apertures (i.e., \texttt{MAG\_APER}) of 0.5"-diameter, and Kron apertures \citep[i.e., \texttt{MAG\_AUTO},][]{Kron_1980} following the same prescription we adopted in \citet[][see section 3.2]{Rinaldi_2022}.  We chose a circular-aperture flux over a Kron flux  when the sources were fainter than a given magnitude. In this case, since we were dealing with very deep images, we decided to consider mag = 27 as our faint limit for the Kron aperture. This final decision has been taken after several tests we performed with the \textit{HST} photometry, comparing our fluxes with the HLF photometric catalogue from \citet{Whitaker_2019}. We corrected the aperture fluxes to the total. For \textit{HST}, these corrections are well known\footnote{Aperture corrections for \href{https://www.stsci.edu/hst/instrumentation/acs/data-analysis/aperture-corrections}{HST/ACS};}\footnote{Aperture corrections for \href{https://www.stsci.edu/hst/instrumentation/wfc3/data-analysis/photometric-calibration/ir-encircled-energy}{HST/WFC3-IR};}\footnote{Aperture corrections for \href{https://www.stsci.edu/hst/instrumentation/wfc3/data-analysis/photometric-calibration/uvis-encircled-energy}{HST/WFC3-UVIS};}. For \textit{JWST}, instead, we estimated the aperture corrections using the software \texttt{WebbPSF}\footnote{The software \texttt{WebbPSF} is available at the following \href{https://webbpsf.readthedocs.io/en/latest/}{link};}.

Moreover, we adopted a minimum error of 0.05~mag for all the HST photometry because \texttt{SExtractor} typically underestimates photometric errors \citep[e.g.,][]{Sonnett_2013}.  We decided to adopt this minimum error value for the \textit{JWST} images as well to account for possible uncertainties in the NIRCam and MIRI flux calibrations.

Finally, all our fluxes have been corrected for Galactic extinction. Those values have been estimated adopting a python package called \texttt{dustmap}\footnote{The \texttt{dustmap} python package is available at the following \href{https://github.com/gregreen/dustmaps/blob/master/docs/index.rst}{link};}. As a sanity check, we compared the correction factors for the \textit{HST} filters with \citet{Schlafly_2011}, finding an excellent agreement with the values we can recover following their prescription, as expected.

\subsection{SED fitting}
We performed the SED fitting and derived the properties of our sources by making use of the code \texttt{LePHARE} \citep{LePhare_2011}. We constructed the libraries for \texttt{LePHARE} by adopting the same configuration we used in \citet[][see section 4]{Rinaldi_2022}.
Briefly, we considered the stellar population synthesis (SPS) models proposed by \citet[][hereafter BC03]{BC_2003}, based on the Chabrier IMF \citep{Chabrier_2003}. We made use of two different star formation histories (SFHs): a standard exponentially declining SFH (known as “$\tau$-model”) and an instantaneous burst adopting a simple stellar population (SSP) model. In particular, we adopted two distinct metallicity values, a solar metallicity (Z$_{\odot}$ = 0.02) and a fifth of solar metallicity (Z = 0.2Z$_{\odot}$ = 0.004). Moreover, to take into account the strong contribution from the nebular emission lines that can occur at very young ages, we also considered \texttt{STARBURST99} templates \citep[][hereafter SB99]{SB99} for young galaxies (age $\leq 10^7 \, \rm yr$) with constant star formation histories. We considered the \citet{Calzetti_2000} reddening law in combination with \citet{Leitherer_2002} to better constrain wavelengths below 912 {\AA}. In particular, we adopted the following color excess values:  $0 \leq E(B-V)\leq 1.5$, with a step of 0.1. We also decided to run \texttt{LePHARE} between $z = 0$ and $z = 20$, by considering the following steps: $\Delta z = 0.04$ between $z = 0$ and $z = 6$ and $\Delta z = 0.1$  between $z = 6$ and $z = 20$ (291 steps in total).   We summarize the parameters we adopted to perform the SED fitting in Table \ref{SED_fitting_parameters}.

\begin{deluxetable*}{ccchlDlc}[ht!]
\tablenum{1}
\tablecaption{The parameters we used to perform the SED fitting with \texttt{LePHARE} by adopting BC03 and \texttt{SB99} models\label{SED_fitting_parameters}.}
\tablewidth{0pt}
\tablehead{
\colhead{Parameter} & \colhead{} & \colhead{}
}
\startdata
Templates & \citet{BC_2003} & \citet{SB99} \\
$e-$folding time ($\tau$) & 0.01 - 15 (8 steps) + SSP & constant SFH \\
Metallicty (Z) & 0.004; 0.02  ($=$Z$_{\odot}$) & 0.008; 0.001\\
Age (Gyr) & 0.001 - 13.5 (49 steps) & 0.001 -  0.1 (6 steps) \\
& \multicolumn{2}{c}{Common values}\\
Extinction laws & \multicolumn{2}{c}{\citet{Calzetti_2000} + \citet{Leitherer_2002}}   \\
E(B-V) & \multicolumn{2}{c}{ 0-1.5 (16 steps)}  \\ 
IMF & \multicolumn{2}{c}{\citet{Chabrier_2003}}    \\ 
Redshift & \multicolumn{2}{c}{0-20 (291 steps)}  \\
Emission lines & \multicolumn{2}{c}{Yes}    \\
Cosmology (H$\mathrm{_{0}}$, $\mathrm{\Omega_{0}}$, $\mathrm{\Lambda_{0}}$) & \multicolumn{2}{c}{70, 0.3, 0.7}
\enddata
\tablecomments{For the run with \texttt{SB99} models, we used the same configuration as for the BC03 models for the extinction law, E(B-V), IMF, redshift interval, and cosmology. Moreover, for the run with \texttt{SB99}, we opted for only 6 steps in age because nebular emission lines only matter for very young ages.}
\end{deluxetable*}

We estimated upper limits for each source that \texttt{SExtractor} was not able to detect. To do that, around each source, we placed random circular apertures (0.5\arcsec-diameter) to estimate the background r.m.s. (1$\sigma$). For \texttt{LePHARE}, we opted to use the 3$\sigma$ upper limit for the flux in those filters where we did not have a detection. Finally, for all those sources for which we did not have any photometric information (e.g., the MIRI/F560W and NIRCam coverage areas are different), we simply ignored those filters during the SED fitting (i.e., we used $-99$ as input flux in \texttt{LePHARE}).

\section{Selection of strong (H$\beta$+[\ion{O}{3}]) and H$\alpha$ emitters at $\lowercase{z} \simeq 7-8$}\label{sec_3}

\texttt{LePHARE} returns the best-fit SED and derived parameters for each source. We performed two different runs with \texttt{LePHARE}, one adopting BC03 models only and the other one adopting \texttt{SB99} models only. Therefore, we created the final catalogue choosing for each source the best $\chi_{\nu}^{2}$ between the BC03 and \texttt{SB99} solutions. Finally, we cleaned our catalogue of possible stars. To do so, we first cross-matched our catalogue with {\em Gaia Data Release 3} \citep[][GAIA DR3]{GAIA_DR3}. Then, we looked at the stellarity parameter (i.e., \texttt{CLASS\_STAR}) we have from \texttt{SExtractor}. In particular, we applied the same criterion adopted in \citet[][Section 3.1]{Caputi_2011}. We removed all those sources that have \texttt{CLASS\_STAR} $>$ 0.8 and occupy the stellar locus in the (F435W $-$ F125W) versus (F125W $-$ F444W) colour$-$colour diagram. In total, less than $\simeq$ 1\% sources have been discarded from our full catalogue because they have been classified as stars (8 of them have been identified in GAIA DR3).

Since our goal is to look for potential (H$\beta$ + [\ion{O}{3}]) and (H$\alpha$+[NII]+[SII]) emitters in the XDF at z $\simeq 7 - 8$, we only focused on those sources for which the best photometric redshift falls in that redshift range.

For each candidate, we created postage stamps to make a careful visual inspection in order to exclude all those galaxies that either fall on stellar spikes or are heavily contaminated by the light of the nearby sources. After this visual inspection,  we were left with 58 robust galaxy candidates at $z\simeq 7-8$.

Among these sources, we searched for (H$\beta$ + [\ion{O}{3}]) and H$\alpha$ emitters.  We first analysed if they show a flux excess in the following three bands: NIRCam/F430M, NIRCam/F444W, and MIRI/F560W. The first two filters have been used to look at the flux enhancement produced by (H$\beta$ + [\ion{O}{3}]). In turn,  MIRI/F560W has been used to look at the flux excess produced by H$\alpha$. 

To convert the flux excess into an EW$_{0}$ we followed the canonical approach described by \citet{Marmol_2016}. Following that procedure, we know that:

\begin{equation}
    \mathrm{EW_{0} = \frac{W_{rec}}{1+z}(10^{(-0.4\Delta mag)} - 1)},
\end{equation}
where W$_{rec}$ is the rectangular width of the filter containing the emission line in question, in our case (H$\beta$ + [\ion{O}{3}]) or H$\alpha$,  and $\Delta$mag is the difference between the observed magnitude in that filter and the synthetic magnitude\footnote{For each galaxy, \texttt{LePHARE} returns the synthetic magnitude in each filter (i.e., $\rm mag_{syn}$) for the best-fit model along with the stellar parameters;} from the SED fitting (i.e., $\Delta$mag = m$\rm_{obs}$ - m$\rm_{syn}$) that we adopted as a proxy for the continuum emission.

Therefore, to estimate the flux excess, we assumed that the continuum flux was well described by the synthetic NIRCam/F460M obtained from the best-fit template for each galaxy. In particular, we selected all those sources for which $\rm |mag_{obs} (F460M) - mag_{syn}(F460M)|\leq 2\; \times$~mag$\rm_{err}$(F460M), where mag$_{\rm obs}$ and mag$_{\rm syn}$ are the observed and best-fit synthetic magnitudes, respectively. This condition ensures that the continuum at 4.6$\mu$m can be considered flat within the error bars. We also double-checked if this condition was satisfied in NIRCam/F480M.

Once we selected all those sources that survive the condition described above, we estimated the flux excess in the following way: $\rm \Delta mag = (mag_{X} - mag_{cont}$), where $\rm mag_{X}$ represents the magnitude in one of the filters we chose to select (H$\beta$+[\ion{O}{3}]) or H$\alpha$, and $\rm mag_{cont}$ refers to $\rm F460M_{syn}$. We highlight that this selection is purely based on the photometric excess we considered above. None of our derivations is based on emission lines modelled by \texttt{LePHARE}. For a conservative approach, we only considered those galaxies for which the flux excess with respect to the stellar continuum satisfies the following condition: $\Delta$mag $<$ $-0.2$.
Note that a  $\Delta$mag~$=-0.2$ in NIRCam/F430M corresponds to a EW$_{0}$~$\simeq 58 \, \rm \AA$ at $z=7$, while in NIRCam/F444W it would imply an EW$_{0}$~$\simeq 270 \, \rm \AA$. For MIRI/F560W, the same $\Delta$mag would correspond to an EW$_{0}$~$\simeq 239 \, \rm \AA$ at the same redshift.

\begin{figure*}[ht!]
    \centering
    \includegraphics[width = \textwidth]{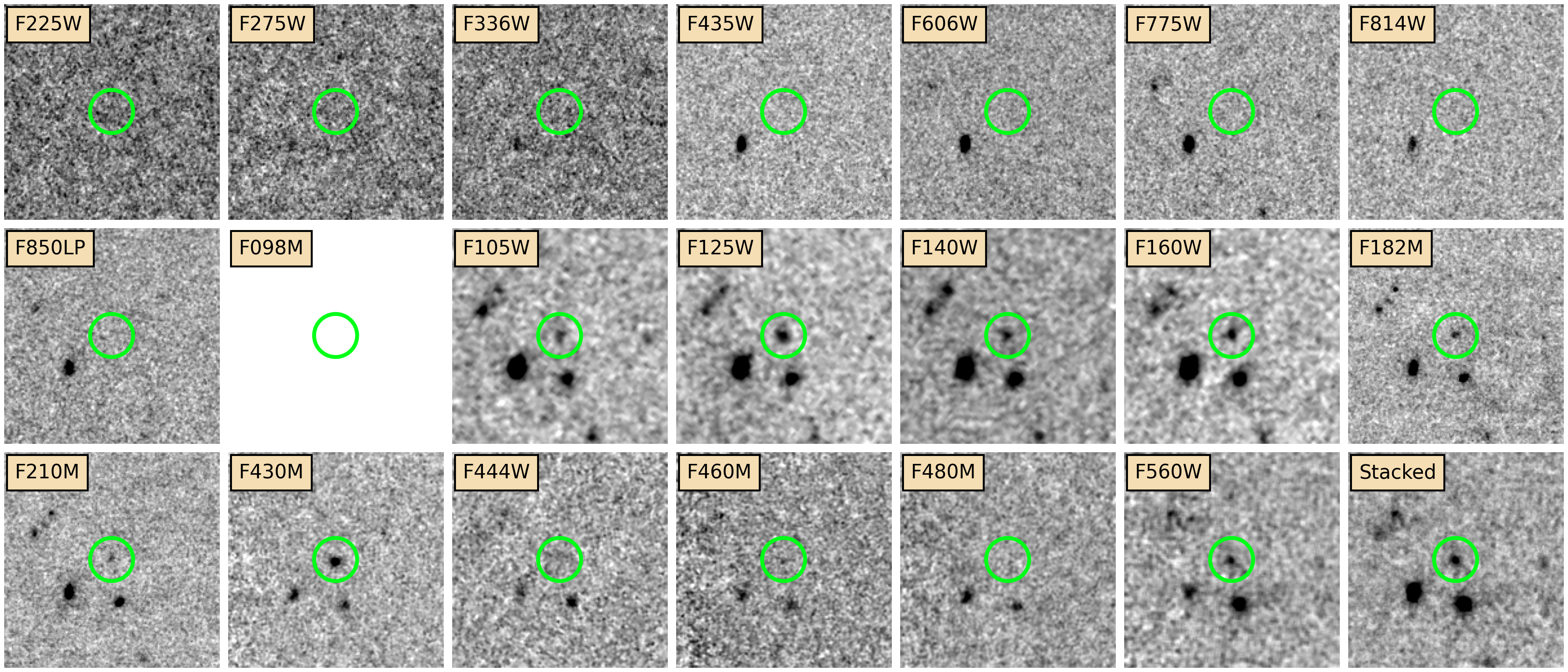}
    \caption{Postage stamps (5" $\times$ 5") of one of our (H$\beta$ + [\ion{O}{3}]) candidates (ID: 9434, z$_{best}$ = 7.68 $_{-0.01}^{+0.03}$). The last postage stamp refers to the stacked image we adopted as the detection map with \texttt{SExtractor}. Here we show all the bands we used in our analysis, from 0.2$\mu$m to 5.6$\mu$m. The green circle has been placed to only guide the eye on the source. In particular, from these postage stamps, there is a clear excess at 4.3$\mu$m. This source shows an excess in MIRI/F560W as well.} 
    \label{Fig_2}
\end{figure*}

We inspected again the postage stamps of the 58 possible candidates, after estimating the flux excess in each band (NIRCam/F430M, NIRCam/F444W, and MIRI/F560W), to make a cross-match between the values we got for $\Delta$mag and the visual inspection of the sources themselves. We also examined the best-fit SED for each galaxy. This safely allowed us to conclude that 18 sources can be securely classified as (H$\beta$ + [\ion{O}{3}]) emitters. These emitters constitute $\simeq 31\%$ of our total galaxy sample at $z\simeq 7-8$ (see Fig. \ref{Fig_2} where we show the multiwavelength images of an example source). The derived EW$_{0}$ values cover a wide range that goes from a minimum of $87.5^{+30}_{-27} \, \rm \AA$ to a maximum value of $2140.4^{+970}_{-154} \, \rm \AA$, with a median $\left \langle \rm EW_{0}  \right \rangle \simeq 943_{-194}^{+737} \, \rm \AA$ (lower and upper errors refer to the 16th and 84th percentiles). This value is higher, but still marginally consistent with the error bars,  than that derived by \citet{Labbe_2013} from \textit{Spitzer Space Telescope} observations of bright $z\simeq 8$ galaxy candidates. Out of the 18 (H$\beta$ + [\ion{O}{3}]) emitters, $83\%$ have a best-fit SED with sub-solar ($\rm 0.2 \, Z_\odot$) metallicity and the remaining $\simeq17\%$ with solar ($\rm Z_\odot$)  metallicity.

Among the 18  (H$\beta$ + [\ion{O}{3}]) emitters at $z\simeq 7-8$, a total of 16 lie on the ultra-deep MIRI $5.6 \, \rm \mu m$ coverage field. Out of them, 12 show a significant $5.6 \, \rm \mu m$ flux excess with respect to the continuum (as defined above), which we interpret as the presence of the (H$\alpha$+[NII]+[SII]) line complex at $z\simeq 7-8$. To obtain the net value of the H$\alpha$ EW$_{0}$, we applied the correction recipes provided by \citet{Fritze_2003}, as follows: $f$(H$\alpha$) = 0.63$f$(H$\alpha$ + [NII] + [SII]) for a solar metallicity, and  $f$(H$\alpha$) = 0.81$f$(H$\alpha$ + [NII] + [SII]) for a $\rm 0.2 \, Z_\odot$ metallicity. Note that with this procedure we are assuming that the stellar and gas metallicities are similar in these galaxies.

\begin{figure*}[ht!]
    \centering
    \includegraphics[width = \textwidth]{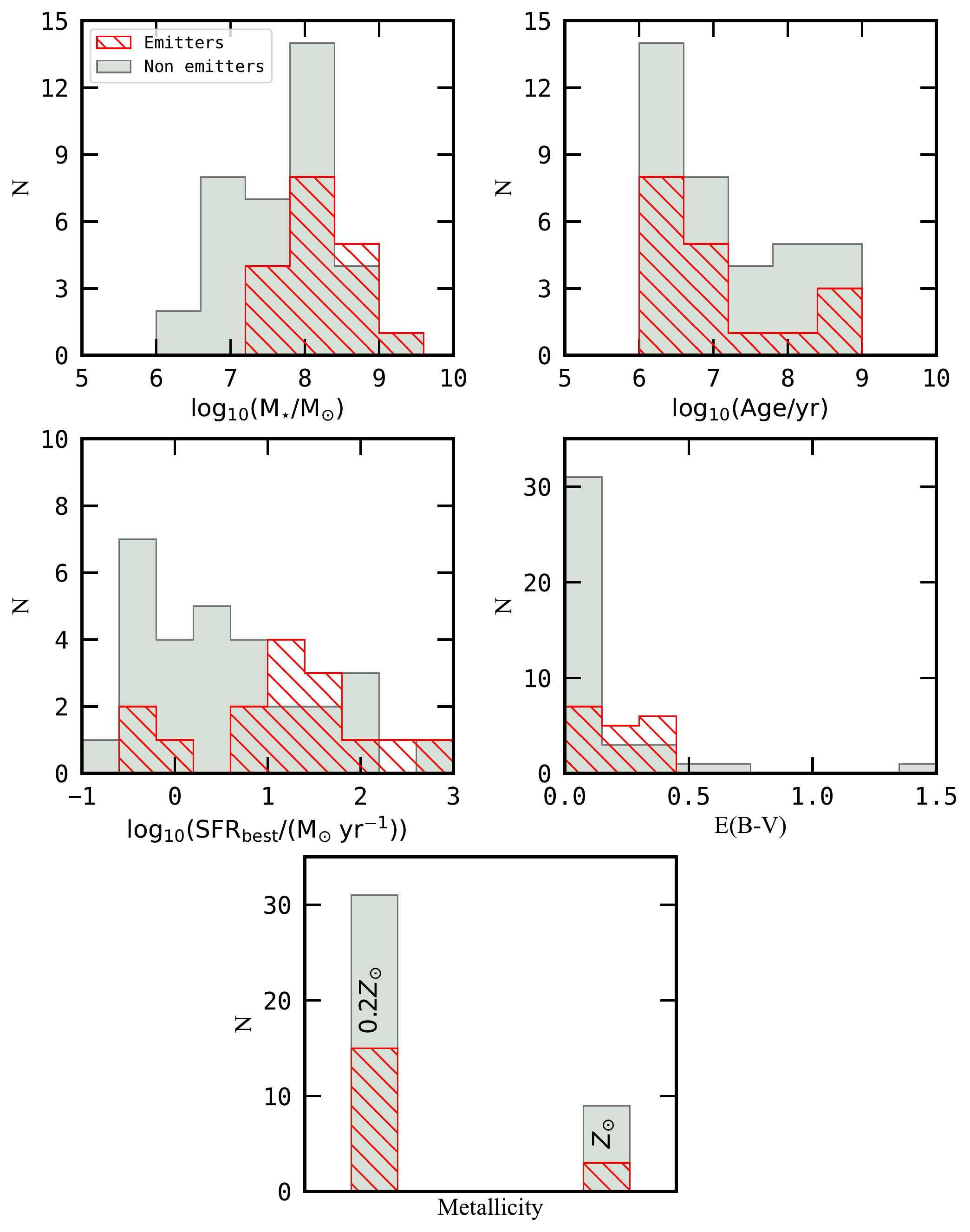}
    \caption{Comparison of the best-fit properties for emitters ((H$\beta$ + [\ion{O}{3}]) and H$\alpha$) and non-emitters at $z=7-8$: stellar mass and age (upper row); star formation rate and colour excess (middle row); and metallicity (bottom row). No significant differences have been noticed between the two populations for most of the stellar parameters, as determined by performing a two-sample Kolmogorov-Smirnov test. Differences in SFR$_{\rm best}$ between emitters and non-emitters might be explained by the fact we are only looking at strong emitters that show a higher SFR$_{\rm best}$.} 
    \label{stellar_parameters}
\end{figure*}

We also compared the derived stellar properties from the SED-fitting between the (H$\beta$ + [\ion{O}{3}]) and H$\alpha$ emitters and non-emitters. Performing the two-sample Kolmogorov-Smirnov test, we do not find any significant difference between the two samples in terms of ages, E(B-V), metallicity, and stellar mass. Regarding the SFR$_{\rm best}$ distributions, we see a difference between the two populations (SFR$_{\rm best}$ for the emitters tend to be higher than SFR$_{\rm best}$ for the non-emitters) that might reflect the fact that we are looking at strong emitters (i.e., SFR is higher). We show these distributions in Fig. \ref{stellar_parameters}.

\section{Results}\label{sec_4}

\begin{figure*}[ht!]
    \centering
    \includegraphics[width = 0.49 \textwidth, height = 0.30 \textheight]{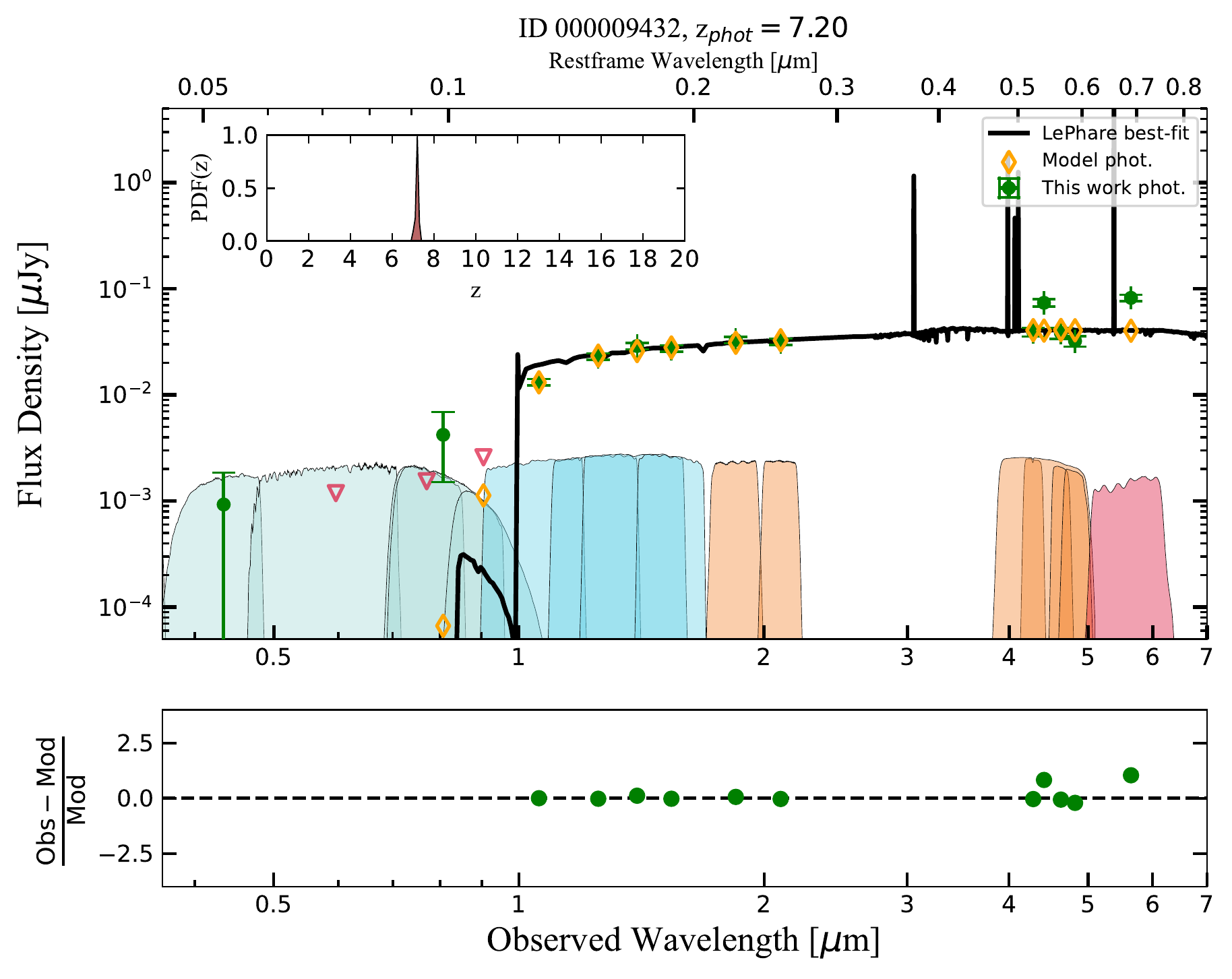}
    \includegraphics[width = 0.49 \textwidth, height = 0.30 \textheight]{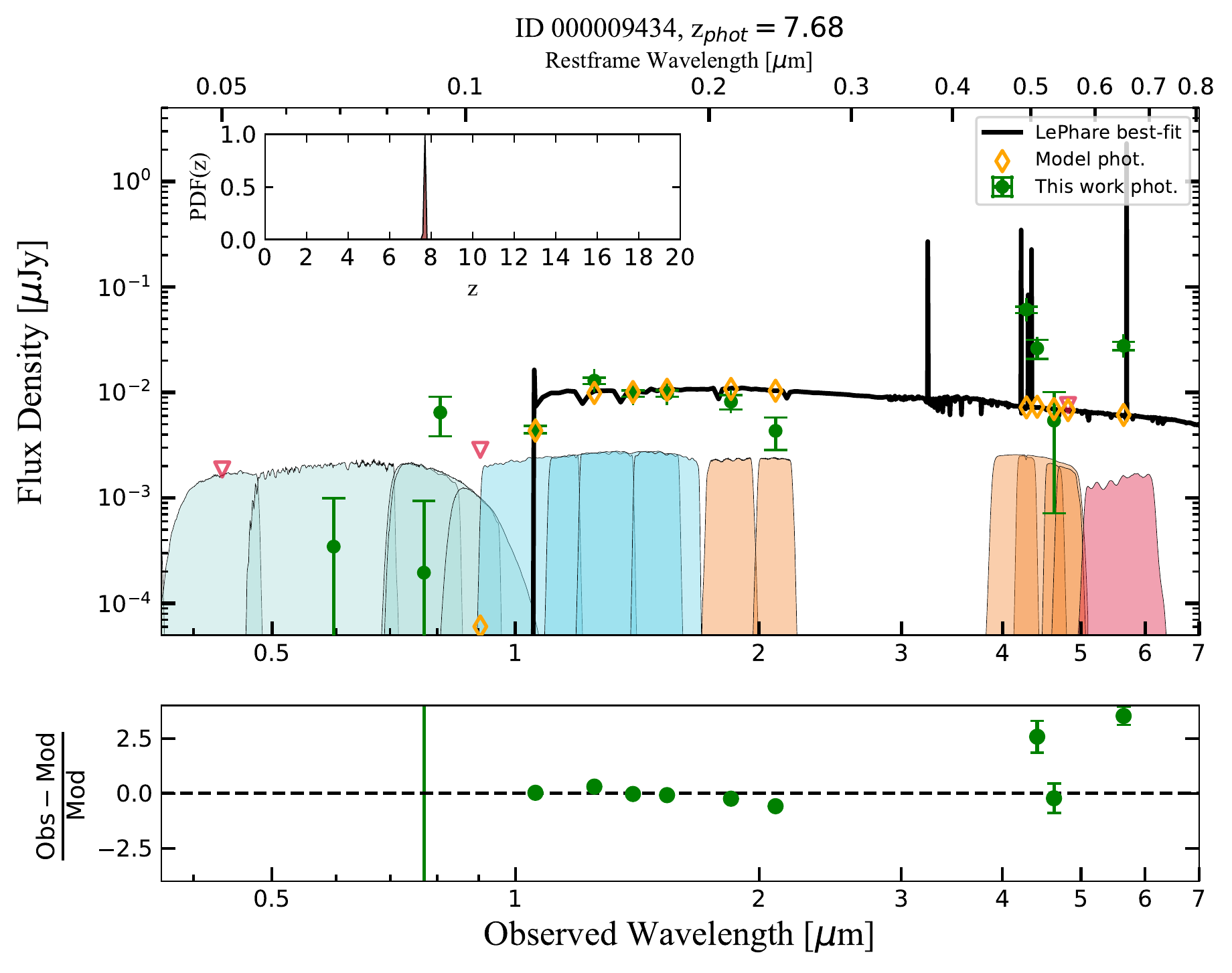}
    
    \caption{The best-fit SEDs for two examples of line-emitter candidates at $z\simeq 7-8$. On the left panel, we show a source at $z_{phot} \simeq 7.19$ (ID = 9432). On the right panel, we show another source at $z_{phot} \simeq 7.68$ (ID = 9434, shown in Fig. \ref{Fig_2}). Both panels show how well-constrained the best-fit SEDs and the derived photometric redshifts are, which is evident by simple inspection of the best-fit templates and their probability density functions (PDF(z)). In each case, we notice the clear presence of an excess in F430M, F444W, and F560W, which we adopted as the criterion to select our sample of (H$\beta$ + [\ion{O}{3}]) and H$\alpha$ emitters.}
    \label{Fig_3}
    
\end{figure*}

Once we estimated the stellar properties of our candidates by performing the SED fitting with \texttt{LePHARE}, we analysed the properties of these sources by comparing our results with the recent literature at high redshifts. Before doing that, we first ensured that the stellar masses we inferred with \texttt{LePHARE} were not affected by the presence of the flux excess we estimated in F430M, F444W, and F560W. To do that, we re-run \texttt{LePHARE} following the
methodology explained by \citet{Caputi_2017}. This time, for each source, we turned off those bands (NIRCam/F430M, NIRCam/F444W, and MIRI/F560W) in which we found a flux excess (i.e., $-99$ following \texttt{LePHARE}'s prescription). Moreover, for this run, we fixed the redshifts adopting the photometric ones we estimated from the original run. Doing this test allows us to ensure that our stellar mass estimates are not affected by any emission line that falls in one of those filters. We found a good agreement within 2$\sigma$. Finally, we also inspected that the stellar continuum was well-described by inspecting the best-fit SEDs we obtained from \texttt{LePHARE}. In Fig. \ref{Fig_3} we show two examples (ID: 9432, 9434) of the best-fit SEDs for the candidates we have in our sample.

\subsection{Emission line EW versus Stellar Mass and Age in galaxies at z $\simeq 7 - 8$}

\begin{figure}[ht!]
    \centering
    \includegraphics[width = 0.49 \textwidth, height = 0.30 \textheight]{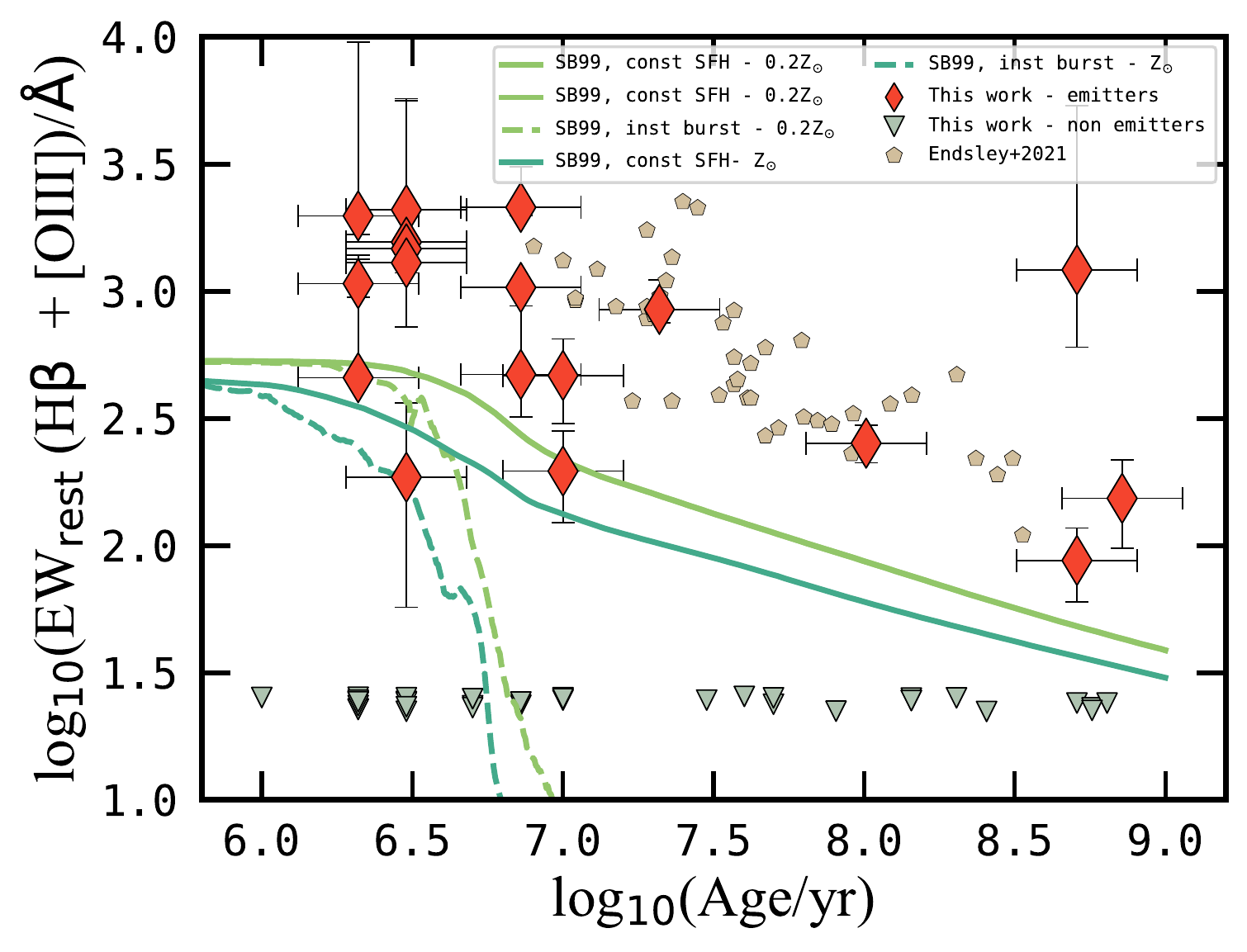}
    \caption{Age versus (H$\beta$ + [\ion{O}{3}]) EW$_{0}$ for our line emitters at $z=7-8$. The grey triangles refer to the EW$_{0}$ upper limits that we estimated for all the  “non-emitter” galaxies in our sample. We also show the data points from \citet{Endsley_2021} to make a comparison with the recent literature at high redshift, albeit in a much higher luminosity regime. The curves refer to the evolution of the H$\beta$ EW$_{0}$ as a function of age expected from SB99 models, corresponding to the two metallicities (solar and sub-solar) that we have considered in our work and for two different SFHs. A clear anti-correlation between EW$_{0}$ and age is evident in this plot, which is in line with previous findings in the literature at lower redshifts \citep{Reddy_2018}.}
    \label{Fig_EWage}
\end{figure}

Having calculated the (H$\beta$ + [\ion{O}{3}]) EW$_{0}$ for the prominent line emitters, we can compare their best-fit SED properties with those of the other $z\simeq 7-8$ galaxies in our sample. In Fig.~\ref{Fig_EWage}, we show the derived (H$\beta$ + [\ion{O}{3}]) EW$_{0}$ versus best-fit age. From this plot, we can see that all except three of the (H$\beta$ + [\ion{O}{3}]) emitters are characterised by young best-fit ages  ($\leq10^{8} \, \rm yr$), which indicates that these objects may be in their first major star-formation episode. The remaining three objects are older ($>10^8 \, \rm yr$), with two having almost the age of the Universe at their redshifts. This fact suggests that these galaxies could be having a rejuvenation episode, as is known to happen at lower redshifts \citep{Rosani_2020}, as it is unlikely that they could have sustained their high instantaneous SFR values for all of their lifetimes.

The grey triangles in Fig.~\ref{Fig_EWage} refer to the EW upper limits that we estimated for all those galaxies at $z\simeq 7-8$ that do not have a significant flux excess in the NIRCam/F430M band. In contrast to the (H$\beta$ + [\ion{O}{3}]) emitters, the non-emitters span different possible ages at those redshifts, without any bias towards young/old ages. 
 
We also compared our results with the recent literature. In particular, \citet{Endsley_2021} studied a sample of 20 rest-frame ultraviolet (UV) bright  (H${\beta}$ + [\ion{O}{3}]) emitters at $z\simeq 6.8 - 7$ that have been selected over a wide sky area (2.7 deg$^{2}$ in total). \citet{Endsley_2021} found this rare population of very strong (H${\beta}$ + [\ion{O}{3}]) emitters with an EW$_{0}$ $>1200$ \AA. The fact that we find similarly high (H${\beta}$ + [\ion{O}{3}]) EWs$_{0}$ among faint galaxies in a much smaller area of the sky indicates that prominent (H${\beta}$ + [\ion{O}{3}]) emitters were much more common at the EoR than what can be inferred from the brightest galaxies.

Finally, the solid and dashed lines in  Fig.~\ref{Fig_EWage} show the expected variation of the H$\beta$ (only) EW$_{0}$ versus age for \texttt{SB99} model galaxies. 
These theoretical tracks are based on a Chabrier IMF with a stellar mass cut-off of $100M_\odot$ and were obtained both for a solar and a subsolar metallicity ($0.2Z_\odot$), each for a single burst and constant SFH. As expected, our data points are located nicely above these curves, following the trend of the models with constant star formation histories, albeit with higher EW$_{0}$, due to the [\ion{O}{3}] contribution.

Over the past decades, the recombination line equivalent widths have been used as proxies for stellar population age in star-forming galaxies. The ratios between the fluxes of the recombination line, which are sensitive to the instantaneous star-formation rates, and the fluxes of the continuum, which are sensitive to the previous average SFR, are indeed what we define as recombination line equivalent widths \citep[][]{Stasinska_1966}. In particular, \citet[][]{Reddy_2018} found a very strong anti-correlation between (H$\beta$ + [\ion{O}{3}]) EW$_{0}$ and young ages at $z\simeq 1.8 - 3.8$, which does not evolve as a function of redshift at that range of cosmic time. By looking at Fig. \ref{Fig_EWage}, we can see that this anti-correlation is evident also at $z \simeq 7 - 8$ where strong (H${\beta}$ + [\ion{O}{3}]) emitters prefer young ages, which is in line with what has been found at lower redshifts. We double-checked this result by estimating the Spearman's rank correlation coefficient, finding that those two quantities anti-correlate (i.e., Spearman's coefficient $\simeq$ $-$0.5) with a \textit{p}-value $\simeq$ 0.03. Therefore, we can conclude that there is evidence of a moderate anti-correlation between age and $\mathrm{EW_{0}}$(H$\beta$ + [\ion{O}{3}]).

\begin{figure}[ht!]
    \centering
    \includegraphics[width = 0.49 \textwidth, height = 0.30 \textheight]{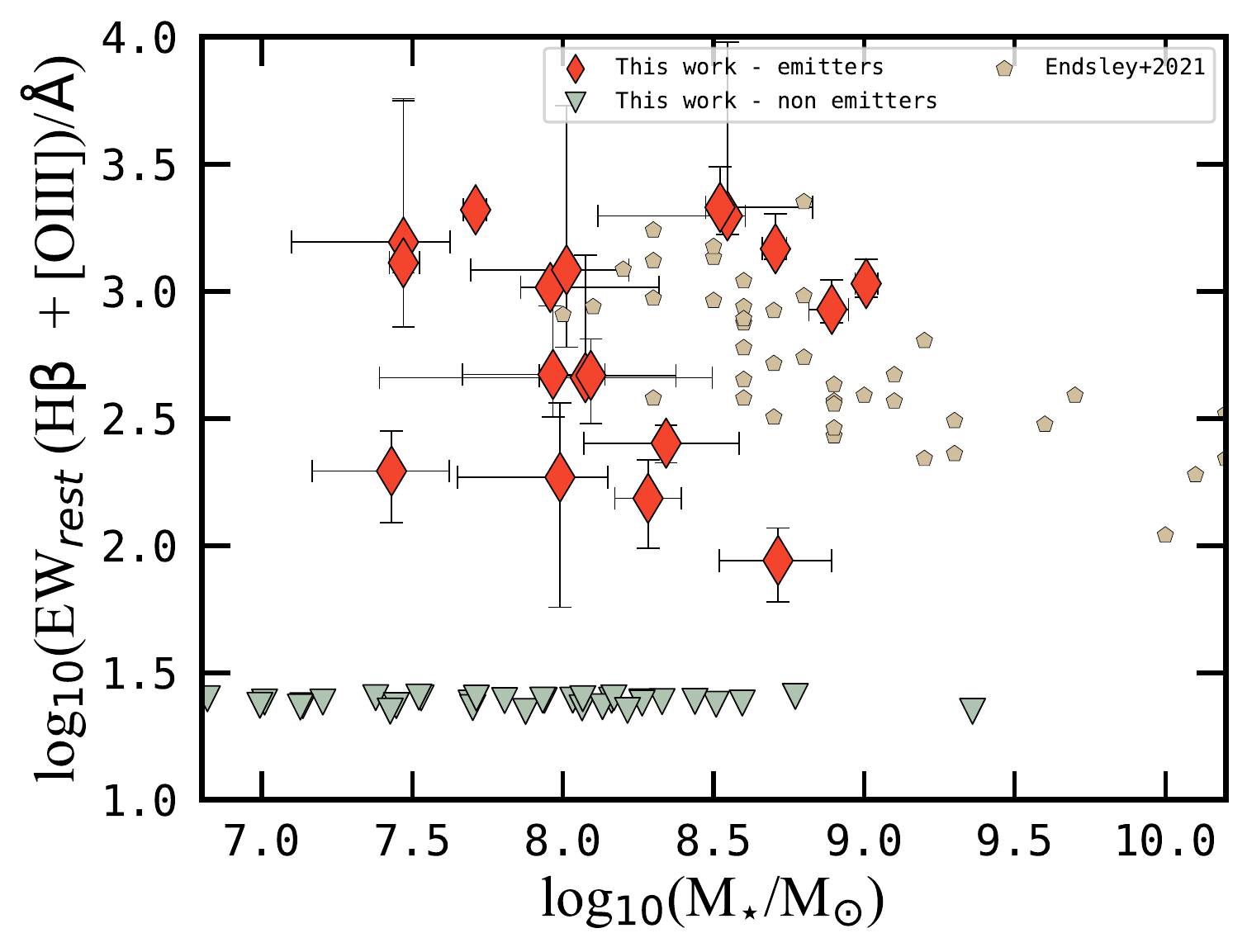}
    \caption{Stellar mass versus (H$\beta$ + [\ion{O}{3}]) EW$_{0}$. The grey triangles refer to the upper limits we estimated for all those galaxies we classified as “non-emitters” during our selection. Also in this case, we report data points from \citet{Endsley_2021} to make a comparison with the recent literature at high redshift. The EW$_{0}$ broadly anti-correlates with stellar mass, similarly to what has been reported by \citet{Reddy_2018} and \citet{Endsley_2021} at lower redshifts.}  
    \label{Fig_EWstm}
\end{figure}

We repeated the same exercise looking, this time, at the derived (H$\beta$ + [\ion{O}{3}]) EW$_{0}$ versus stellar mass for our (H$\beta$ + [\ion{O}{3}]) emitters (Fig. \ref{Fig_EWstm}). Also, in this case, the stellar masses come directly from the best-fit SED obtained with \texttt{LePHARE}. As we can see from Fig. \ref{Fig_EWstm}, our (H$\beta$ + [\ion{O}{3}]) emitters have a stellar mass that ranges from a minimum value of log$_{10}$(M$_{\star}$/M$_{\odot}$) $\simeq$ 7.5 to a maximum value of log$_{10}$(M$_{\star}$/M$_{\odot}$) $\simeq$ 9. In previous works, it has been shown that the normalization of the (H$\beta$ + [\ion{O}{3}]) EW$_{0}$ versus stellar mass relation should increase with redshift \citep[e.g.,][]{Reddy_2018}. Here we find a broad anti-correlation between the two quantities. The grey triangles in Fig. \ref{Fig_EWstm} refer to the upper limits that we estimated for the (H$\beta$ + [\ion{O}{3}]) EW$_{0}$ for the $z \simeq 7-8$ galaxies that are not classified as emitters from a NIRCam flux excess.

\begin{figure}[ht!]
    \centering
    \includegraphics[width = 0.49 \textwidth, height = 0.30 \textheight]{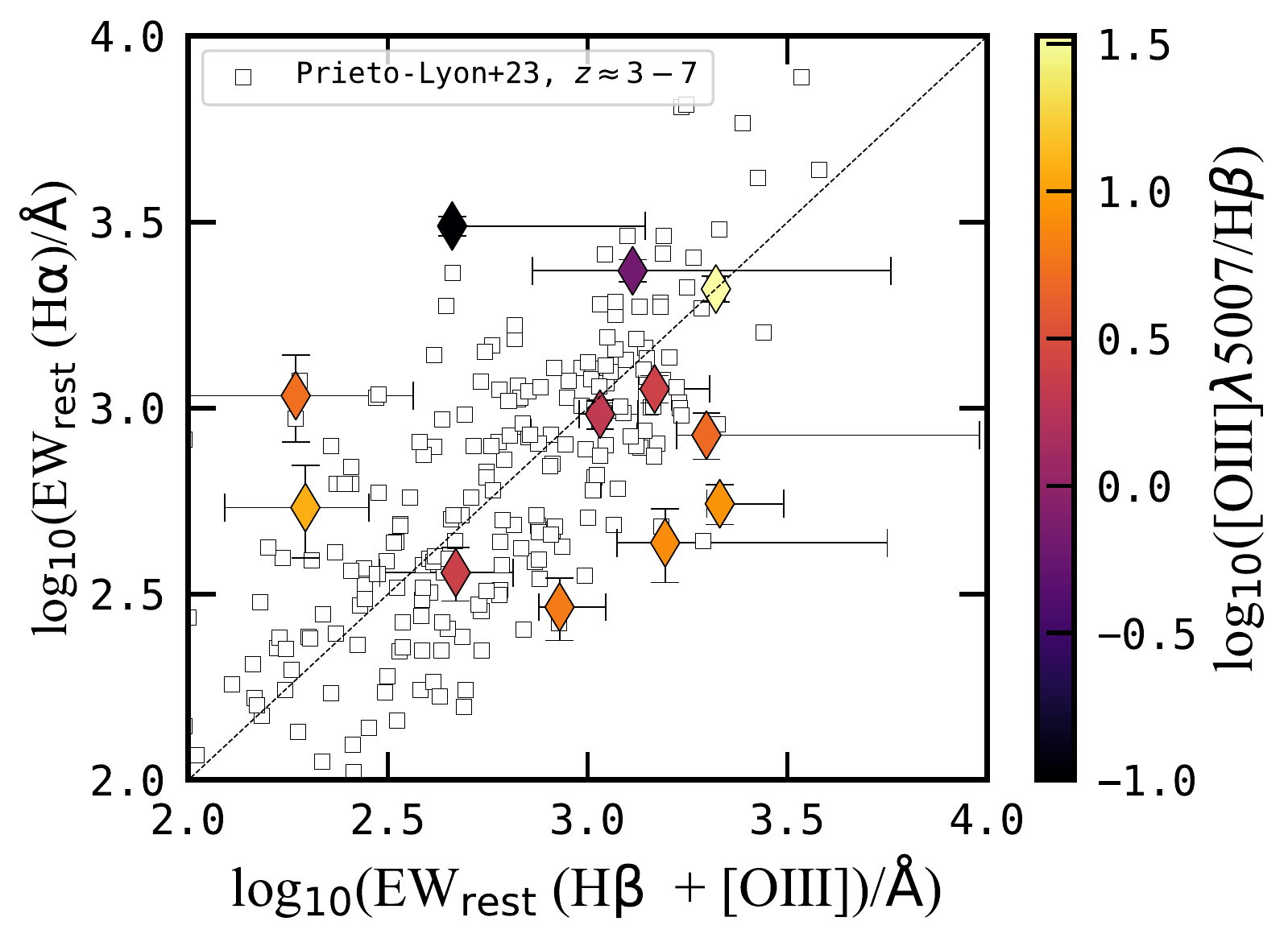}
    \caption{(H$\beta$ + [\ion{O}{3}]) EW$_{0}$ versus H$\alpha$ EW$_{0}$. Here we colour-coded our data points for the [\ion{O}{3}]$\lambda$5007/H$\beta$ flux ratio. Most galaxies have [\ion{O}{3}]$\lambda$5007/H$\beta$ $>$ 1 and they mostly lie on or below the identity line. Among the four galaxies that lie above the identity line, two have [\ion{O}{3}]$\lambda$5007/H$\beta$ $<$ 1, i.e., these line ratios are dominated by H$\beta$. Instead, the other two galaxies above the identity line have [\ion{O}{3}]$\lambda$5007/H$\beta$ $>$ 1 and correspond to cases with non-negligible dust extinction.}
    \label{Fig_6}
\end{figure}

Finally, for all those galaxies that show an ``H$\alpha$ excess'', we compare their (H$\beta$ + [\ion{O}{3}]) EW$_{0}$ versus H$\alpha$ EW$_{0}$, where the ``H$\alpha$ excess'' has been corrected to only take into account the real H$\alpha$ flux following \citet{Fritze_2003}. We show this comparison in Fig.~\ref{Fig_6}. In particular, we also plot the recent results from \citet{Lyon_2023} where they inferred those quantities studying a sample of galaxies at $z\approx 3-7$. We see that our sample is in good agreement with the expected correlation that has been found in \citet{Lyon_2023} as well. As a matter of fact, as we can derive the H$\alpha$ line flux from our data, we can also infer the H$\beta$ line flux independently and separate the contributions of  H$\beta$  and [\ion{O}{3}] for each galaxy, considering the following:

\begin{equation}
f(H\beta) = f(H\alpha) \times  10^{-0.4 \times  1.27 E(B-V)} / 2.86,
\end{equation}

\noindent where $f(\rm H\alpha)$ and $f(\rm H\beta)$ refer to the observed fluxes and $E(B-V)$  is the colour excess obtained from the best-fit SED model. The denominator 2.86 corresponds to assuming case B recombination \cite[e.g.][]{Osterbrock_2006}, while the factor $-1.27=k(\rm H\alpha) - k(\rm H\beta)$ is obtained from the \citet{Calzetti_2000} reddening law.

Once we know the H$\beta$ flux for each source,  we can independently work out the [\ion{O}{3}]$\lambda \lambda$4959, 5007 fluxes for all those emitters that show an H$\alpha$ excess.

The data points in Fig.~\ref{Fig_6} are colour-coded according to each galaxy's [\ion{O}{3}]$\lambda$5007/H$\beta$ ratio. From that figure, we see that most line emitters have [\ion{O}{3}]/H$\beta > 1$, indicating the predominance of [\ion{O}{3}], which is consistent with recent literature findings at similar redshifts. Instead, two galaxies have [\ion{O}{3}]/H$\beta < 1$, i.e. the H$\beta$ line flux is larger than the [\ion{O}{3}] line flux for them. These two galaxies are well above the identity line in Fig.~\ref{Fig_6}, as expected. We separate the H$\beta$ and [\ion{O}{3}]$\lambda$5007 line fluxes simply assuming the case-B recombination  H$\alpha$/H$\beta$=2.86 ratio and the corresponding colour excess mentioned above. The [\ion{O}{3}]/H$\beta < 1$ values could indicate very low metallicities, but this would need to be confirmed with a spectroscopy follow-up of these sources.

\subsection{H$\alpha$-derived SFR and the location of galaxies on the SFR-M$_{\star}$ plane}

For all the 12 H$\alpha$ emitters at $z\simeq 7-8$  as determined from the MIRI $5.6 \, \rm \mu m$ imaging, we estimated their star formation rate (SFR) from their inferred H$\alpha$  luminosities.

After we obtained the net observed H$\alpha$ flux for each source, we converted those fluxes into the intrinsic ones 
by simply applying the Calzetti reddening law. We then estimate the luminosity for the H$\alpha$ emission line and apply the following formula from \citet{Kennicut_1998} to obtain the corresponding SFR(H$\alpha$):

\begin{equation}
     \mathrm{SFR (M_{\odot}\;yr^{-1})} = 7.9\times 10^{-42} \; \mathrm{L_{H\alpha}} \mathrm{(erg} \; \mathrm{s ^{-1}).}
     \label{K98}
\end{equation}

Since the aforementioned formula has been originally calibrated for a Salpeter IMF over $\rm (0.1 - 100)\; M_{\odot}$ \citep{Salpeter_1955}, we applied a conversion factor \citep[][i.e. 1.55]{Madau_2014} to rescale it to a Chabrier IMF \citep[][]{Chabrier_2003}.

\begin{figure}[ht!]
    \centering
    \includegraphics[width = 0.49 \textwidth, height = 0.30 \textheight]{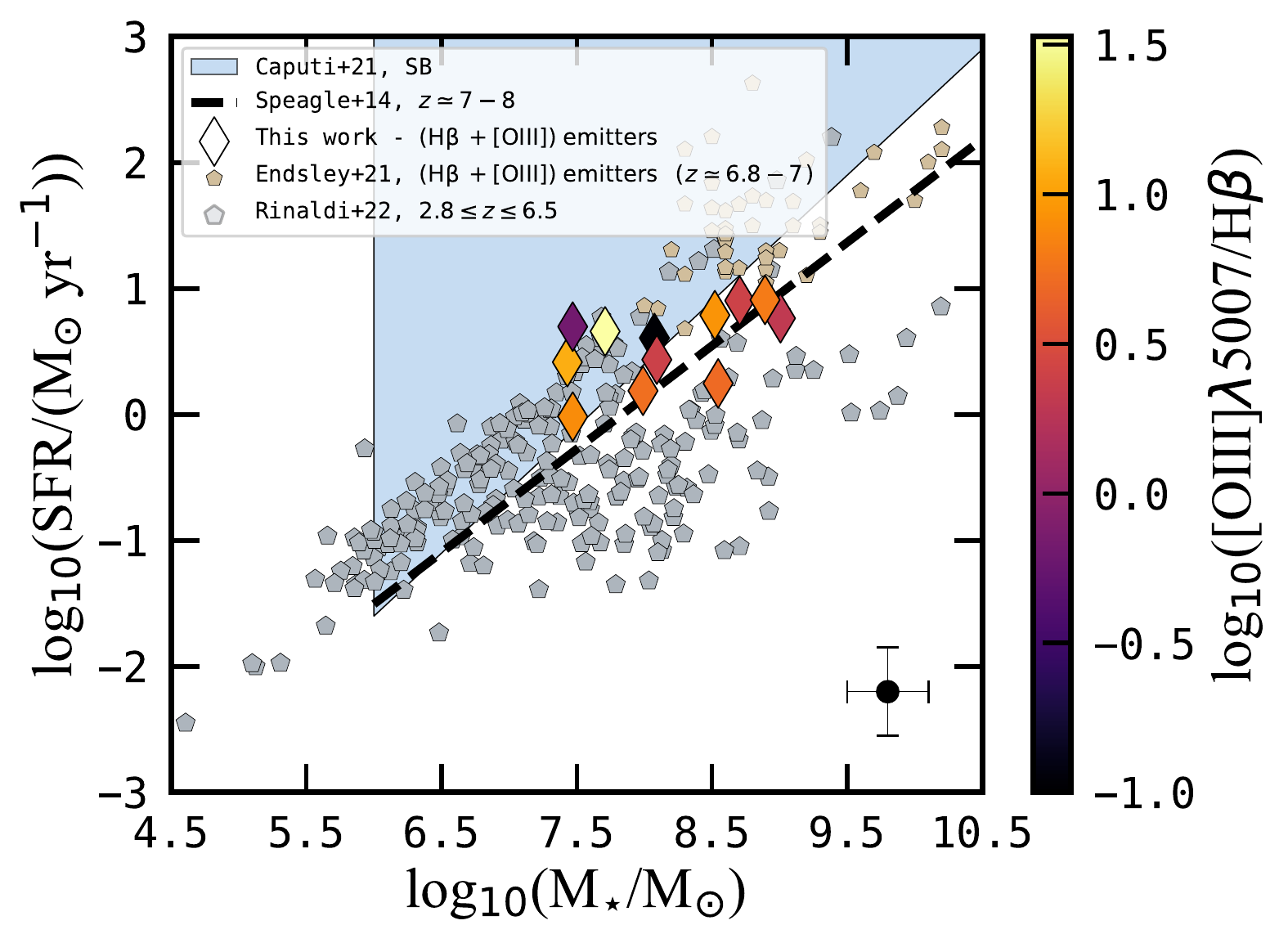}
    \caption{Stellar mass versus SFR. Here we show the SFR$-$M$_{\star}$ plane populated by the SFR directly inferred from the “H$\alpha$ excess”. To make a comparison with the recent literature at high redshifts, we plot data points from \citet{Rinaldi_2022} that give us the opportunity to populate this plane with very low-mass galaxies at $z\simeq2.8-6.5$. We also show data points from \citet{Endsley_2021}, who studied a sample of 20 bright (H$\beta$ + [\ion{O}{3}]) emitters at $z\simeq6.8-7$, and indicate the starburst zone, as defined by \citet{Caputi_2017, Caputi_2021}. We also plot the expected main-sequence of galaxies at $z\simeq 7-8$ from \citet{Speagle_2014}. Our data points are colour-coded by their [\ion{O}{3}]/H$\beta$ ratio. We see no correlation between this ratio and the position of sources on the SFR$-$M$_{\star}$ plane.}
    \label{sfr_plane}
\end{figure}

We then placed our sources on the SFR$-$M$_{\star}$ plane, as we show in Fig.~\ref{sfr_plane}. To make a comparison with the recent literature, we also populated this plane with star-forming galaxies at $z\simeq3.0-6.5$ from \citet{Rinaldi_2022} and (H$\beta$+[\ion{O}{3}])  emitters at $z\simeq6.8-7$ \citep{Endsley_2021}. We also indicate the starburst (SB) zone as determined in \citet{Caputi_2017, Caputi_2021}, which empirically defined as starburst galaxies all those sources with a specific SFR (sSFR)  $\mathrm{> 10^{-7.60}\;yr^{-1}}$. 

We see that 5 ($\simeq 42$\%) of the galaxies that show an ``H$\alpha$ excess'' lie in the starburst zone, while only two are located on the star-formation main sequence \citep[MS; ][]{Brinchmann_2004, Noeske_2007, Peng_2010, Speagle_2014, Rinaldi_2022}. The remaining 5 galaxies appear close, but slightly below the starburst envelope,  in what has been defined in \citet{Caputi_2017} as the star formation valley (SFV), i.e. in between the starburst cloud and the MS, suggesting that they are on the way to/from a starbursting phase. The fact that the vast majority of emitters are in or close to the starburst zone is consistent with the findings of \citet{Endsley_2021} for brighter galaxies, as it can be seen in Fig.~\ref{sfr_plane}.

We also colour-coded our H$\alpha$ emitters according to their [\ion{O}{3}]$\lambda 5007$/H$\beta$ ratios. We find no correlation between these ratios and the position of galaxies on the SFR$-$M$_{\star}$ plane.

\begin{figure}[ht!]
    \centering
    \includegraphics[width = 0.49 \textwidth, height = 0.6 \textheight]{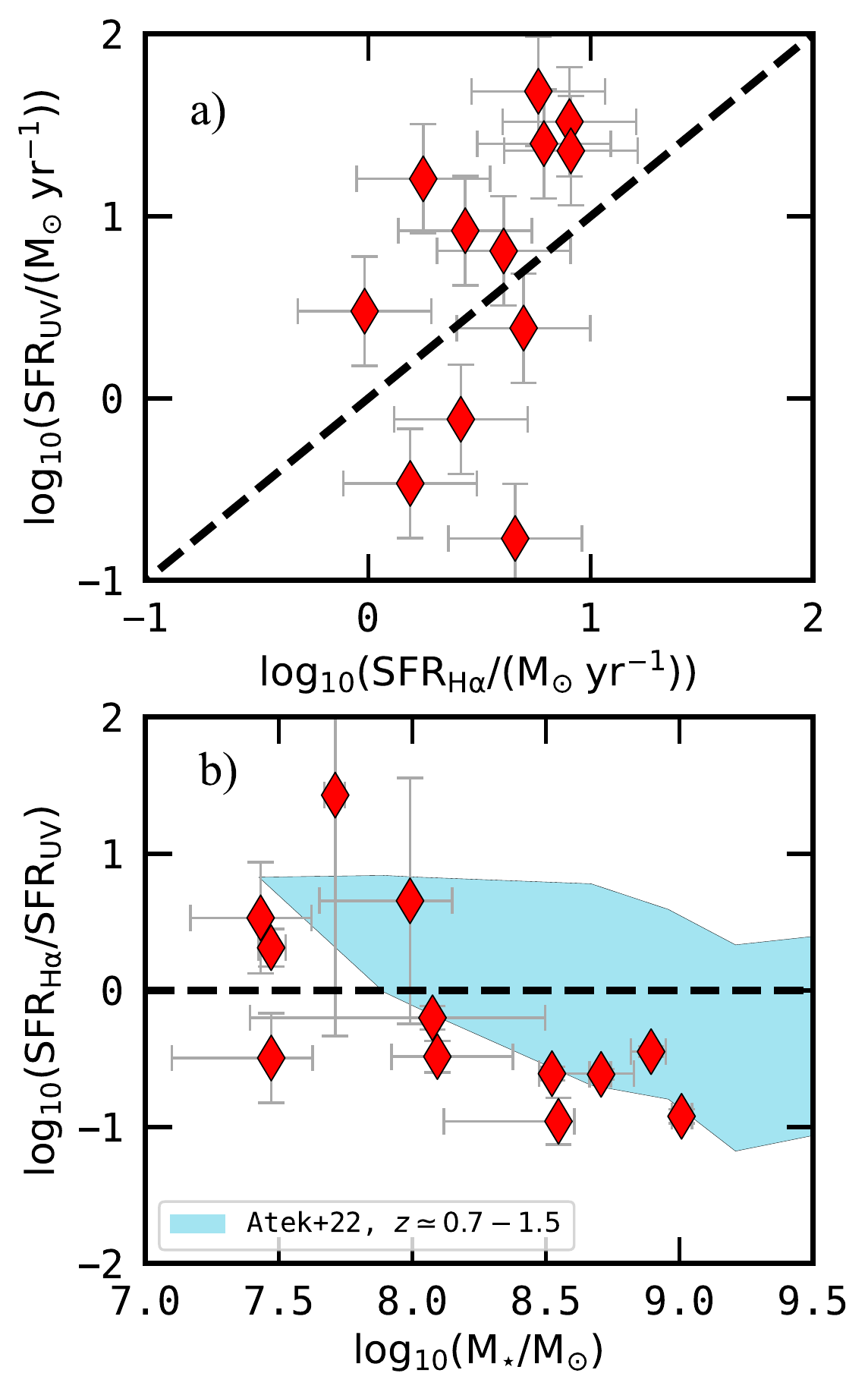}
    \caption{a) Comparison between $\mathrm{SFR_{UV}}$ and $\mathrm{SFR_{H_\alpha}}$. Error bars reflect the usual scatter that has been observed with the Kennicutt's relations we used to derive those two quantities. b) The ratio of $\mathrm{SFR_{H_\alpha}}$ and $\mathrm{SFR_{UV}}$ as a function of stellar mass. Both SFRs have been corrected by adopting the same reddening curve \citep[][]{Calzetti_2000}. The horizontal line indicates a one-to-one ratio. The pale blue shade refers to \citet{Atek_2022} results at lower redshifts.}
    \label{sfr_comparison}
\end{figure}

For the H$\alpha$ sample in Fig. \hyperref[sfr_comparison]{\ref*{sfr_comparison}a} we show the comparison between the two different SFR indicators that we considered in this paper (UV and H$\alpha$ luminosities). From that plot, we clearly see differences between those two indicators ($\mathrm{SFR_{H\alpha}}$ and $\mathrm{SFR_{UV}}$). This finding is not surprising as it has been already pointed out in the literature \citep[e.g., ][]{Velazquez_2021, Atek_2022, Patel_2023}.

Differences between these two SFR tracers (Fig. \hyperref[sfr_comparison]{\ref*{sfr_comparison}a}) may be partly explained by uncertainties in the dust-extinction correction, which mostly affect the UV continuum fluxes, and by our assumption that the dust extinction of the continuum and emission lines is the same and only depends on wavelength.
However, part of the scatter observed in the $\rm SFR_{H\alpha}$ and $\rm SFR_{UV}$ plane may be real and due to the following:
\begin{itemize}
 \item In very young galaxies (below 100 Myr), $\mathrm{SFR_{UV}}$ underestimates the real value because the UV luminosity associated with star formation is still growing. Indeed, when comparing $\rm SFR_{H\alpha}$ and $\rm SFR_{UV}$, one has to take into account age effects as well. UV traces typically 1500-2000{\AA}  (i.e., non-ionizing photons), while H$\alpha$ traces directly $<$ 912{\AA} photons. For example, UV-bright regions without H$\alpha$ emission trace the presence of star-forming clumps dominated by B-type stars and where most massive O-type have already evolved;

 \item Different ionizing photon production efficiencies \citep[e.g.,][]{Nanayakkara_2020, Endsley_2022, Rinaldi_23b}.
\end{itemize}

Finally, by exploiting the FIRE simulations \citep{Hopkins_2014}, \citet{Sparre_2017} showed that the H$\alpha$ measurement of the SFR over a short timescale can fluctuate significantly, up to a factor of ten, compared to the UV indicator.

Following  \citet{Atek_2022} procedure at lower redshifts, in Fig. \hyperref[sfr_comparison]{\ref*{sfr_comparison}b} we inspected the ratio between SFR$\rm_{H\alpha}$ and SFR$\rm_{UV}$ as a function of the stellar mass. We find similar results as \citet[][see their Fig. 8]{Atek_2022} where the ratio of SFR$_{\rm H\alpha}$/SFR$_{\rm UV}$ seems to be generally higher for the low-mass galaxies. Similarly, \citet{Faisst_2019} find that more than 50\% of their sample has SFR$_{\rm H\alpha}$ in excess compared to SFR$_{\rm UV}$, particularly in low-mass galaxies. However, there are still uncertainties in determining the ratio of SFR$_{\rm H\alpha}$/SFR$_{\rm UV}$ and how it changes with different galaxy parameters. As we know from the literature, the SFR indicators use conversion factors from H$\alpha$ and UV luminosities, which assume that the star formation rate is constant. Nonetheless, this assumption may not be that accurate for different SFHs, especially when we consider cases of bursty star formation.

\subsection{The role of the H$\alpha$ emitters in the Cosmic Star Formation History at $z \simeq 7-8$ }

With the SFR values derived in the previous section,  we computed the contribution of the prominent H$\alpha$ emitters to the cosmic Star Formation Rate Density (SFRD) at $z\simeq 7-8$. To do that, we sum up the individual SFRs ($\rm SFR_{H\alpha, total} \simeq 51.39\;M_{\odot}\;yr^{-1}$) and then divide the total by the comoving volume\footnote{We estimated the comoving volume for the entire sky at $z\simeq 7-8$ by using the Cosmology calculator at the following \href{http://www.astro.ucla.edu/~wright/CosmoCalc.html}{link};} encompassed by the area ($\rm A\simeq 4.7\; arcmin^{2}$) and redshift bin (i.e., $z\simeq7-8$) analysed in this work ($\rm V_{sky}\simeq 11580.26\;Mpc^{3}$). We obtain that, at these redshifts, the H$\alpha$ emitters make for $\mathrm{log_{10} (\rho_{SFR_{H\alpha}}/(M_{\odot}\;yr^{-1}\;Mpc^{-3})) \simeq -2.35 \pm 0.3}$.

\begin{figure*}[ht!]
    \centering
    \includegraphics[width = \textwidth]{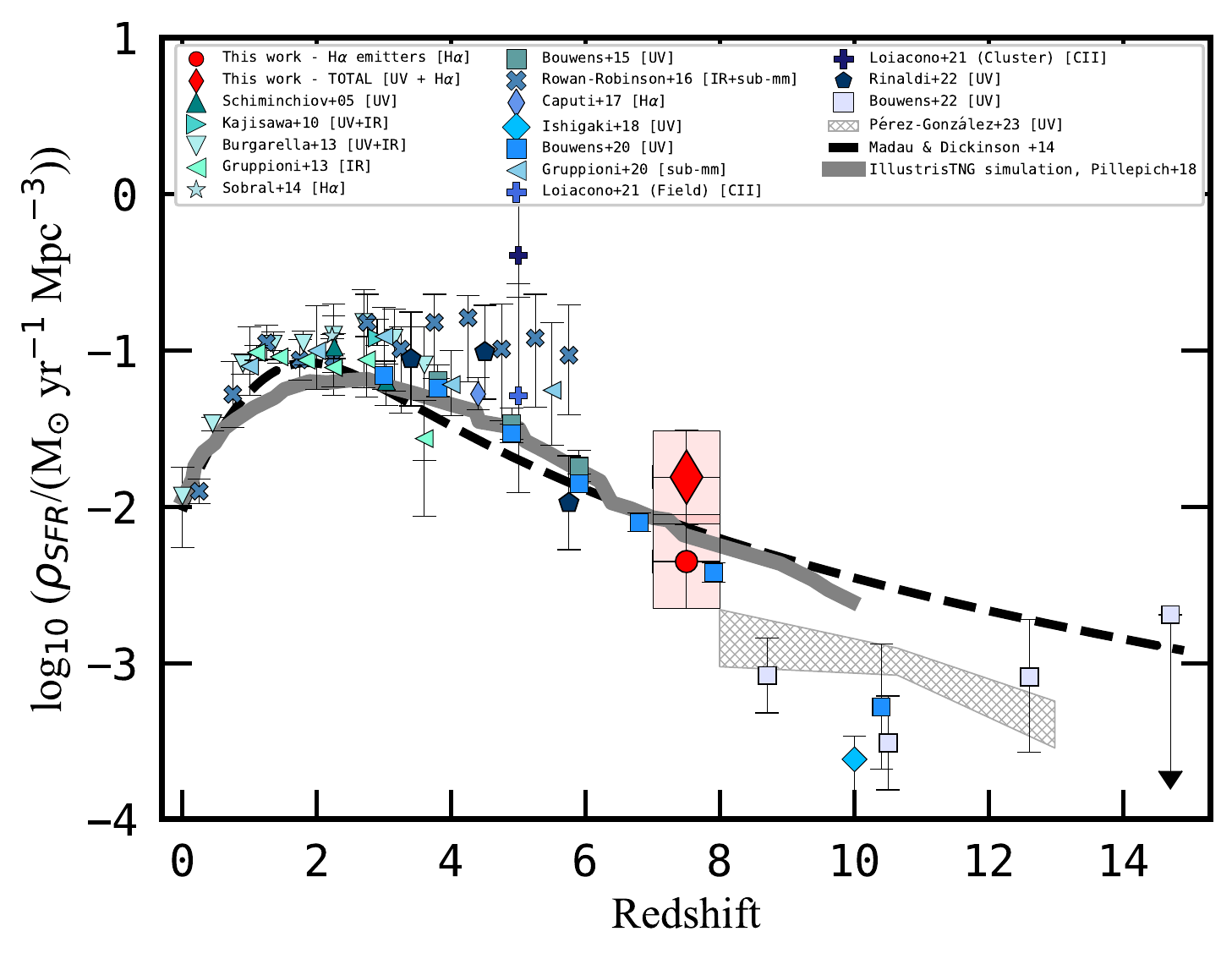}
    \caption{Cosmic star formation rate density as a function of the redshift. The large red circle at $\mathrm{log_{10} (\rho_{SFR_{H\alpha}}/(M_{\odot}\;yr^{-1}\;Mpc^{-3})) \simeq -2.35}$ indicates our estimate at $z \simeq 7-8$, which only takes into account the prominent H$\alpha$ emitters, i.e., it should be considered a lower limit to the real SFRD value at these redshifts. The red diamond at $\mathrm{log_{10}(\rho_{SFR_{tot}}/(M_{\odot}\;yr^{-1}\; Mpc^{-3})) \simeq -1.76}$, instead, refers to the total SFRD we estimated taking into account both the H$\alpha$ emitters and non-emitters at $z\simeq7-8$. For the non-emitters, the SFR directly comes from the UV continuum emission. Other symbols refer to the recent SFRD determinations from the literature, based on different SFR tracers
    \citep[][]{Schiminovich_2005, Kajisawa_2010, Burgarella_2013, Gruppioni_2013, Sobral_14, Bouwens_2015, Rowan_Robinson_2016, Caputi_2017,  Ishigaki_2018, Bouwens_2020, Gruppioni_2020, Loiacono_2021, Rinaldi_2022, Bouwens_2022, Pablo_2023}. The different curves correspond to theoretical predictions. Dashed line: \citet{Madau_2014}. Solid line: \citet{Pillepich_2018}.  All the SFRD values in this figure correspond to a \citet{Chabrier_2003} IMF.
    }
    \label{sfrd}
\end{figure*}

In Fig. \ref{sfrd} we show the redshift evolution of the SFRD as proposed by \citet{Lilly_1996} and \citet{Madau_1996}, the so-called “Lilly-Madau diagram”. In this plot, we show our own estimation of the SFRD, along with a compilation of recent results from the literature based on different SFR tracers. In particular, we also show the SFRD values that have recently been obtained by \citet{Bouwens_2022} using JWST data,  tracing the SFR directly from the UV continuum emission at $z \simeq 9$  to $z \simeq 15$ as well as \citet{Pablo_2023} results at $z\simeq8-13$ from ultra-deep NIRCam images in HUDF-P2 (PID proposal: 1283, PI: G\"oran \"Ostlin). Our inferred SFRD appears to be in good agreement with what has been found in the literature at similar redshifts. We also find a very good agreement with the predictions from theoretical models  \citep[e.g., IllustrisTNG; ][]{Springel2018}. In particular, in Fig. \ref{sfrd} we also show the total SFRD, which has been estimated from both H$\alpha$ emitters and non-emitters at $z\simeq 7-8$ ($\mathrm{log_{10}(\rho_{SFR_{tot}}/(M_{\odot}\;yr^{-1}\;Mpc^{-3})) \simeq -1.76 \pm 0.3}$)\footnote{For the non-emitters, $\rm A  = 5.25\;arcmin^{2}$ (corresponding to the NIRCam coverage) and $\rm V_{sky} = 12999.90\; Mpc^{3}$}. For the non-emitters, the SFR has been obtained from the rest-frame UV continuum luminosity at $2000 \, \rm \AA$ and adopting the conversion formula from \citet{Kennicut_1998}.

\subsection{The evolution of the rest-frame EW(H$\alpha$) as a function of redshift}

\begin{figure}[ht!]
    \centering
    \includegraphics[width = 0.49 \textwidth, height = 0.30 \textheight]{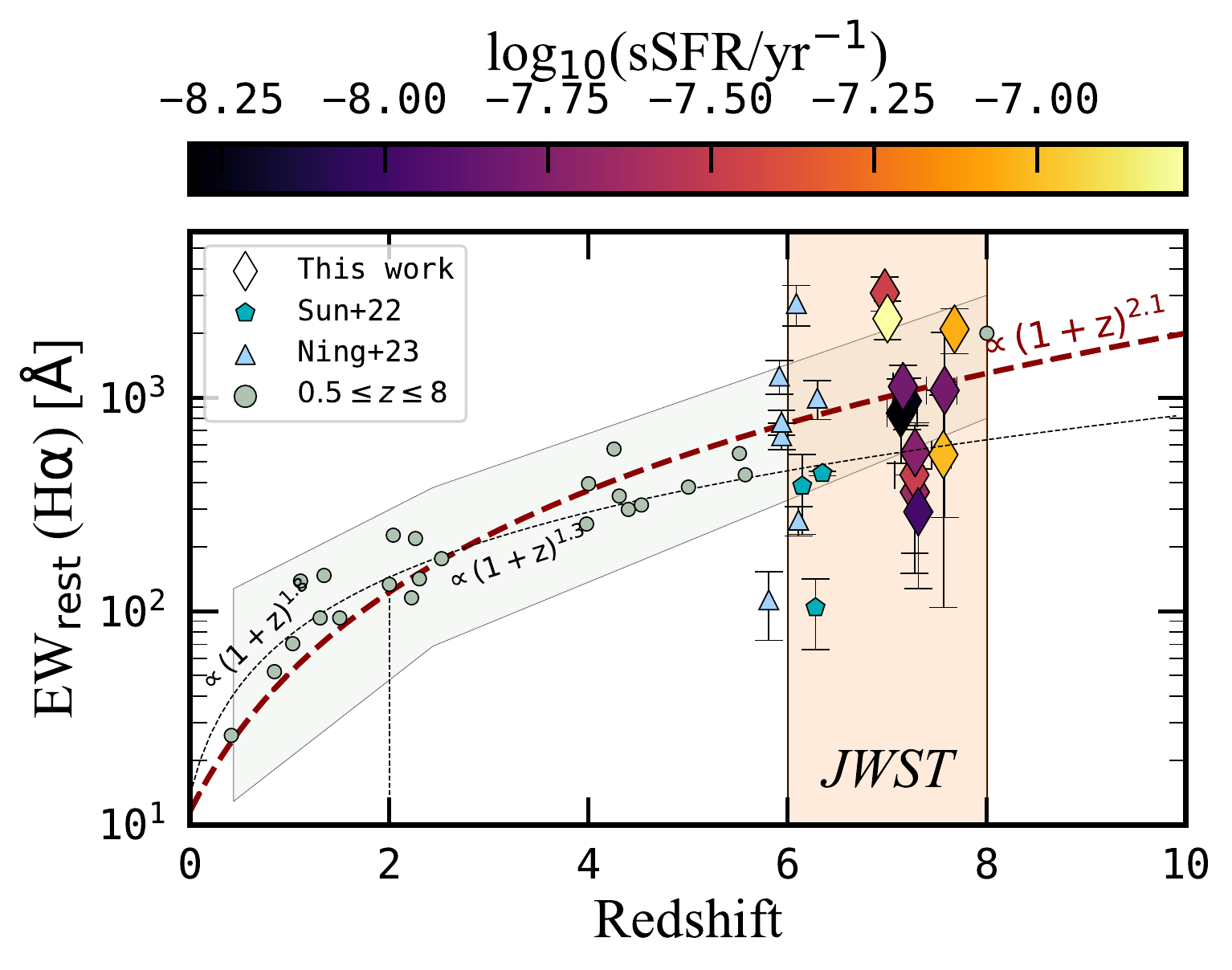}
    \caption{Evolution of the EW$_{0}$(H$\alpha$) as a function of redshift. After re-fitting all the data points, including our own, we find that the  EW$_{0}$(H$\alpha$)  evolution can be described by a single law: EW$\mathrm{_{0}(H\alpha) \propto (1+z)^{2.1}}$  (bold, dark red, and dashed line.) Our data points are colour-coded for sSFR. We also report the recent literature regarding the evolution of the EW$_{0}$(H$\alpha$) as a function of the redshift (tiny, black, and dashed line). The grey shade represents a median estimate of the error bars of the data points from the literature. The orange shade represents the redshift window where JWST is starting to detect these kinds of sources more systematically \citep[e.g.,][]{Boyett_2022, Sun_2022, Yuanhang_2023}. Note that the data point at $z \simeq 8$ from \citet{Stefanon_2022} has been obtained by median stacking a sample of 102 Lyman-break galaxies in the \textit{Spitzer}/IRAC bands from 3.6$\mu$m to 8$\mu$m.
    \label{ew_z}}
\end{figure}

Finally, our derived values of the H$\alpha$ EW$_{0}$ allow us to extend the study of the redshift evolution of this parameter to $z\simeq 7-8$.  In Fig.~\ref{ew_z}  we present our results along with the most recent determinations from the literature  \citep[for sources at $z \simeq 0.5 - 6$,][]{Erb_2006, Shim_2011, Fumagalli_2012, Stark_2013, Sobral_2014, Marmol_2016, Smit_2016, Reddy_2018, Lam_2019, Atek_2022, Boyett_2022, Sun_2022, Yuanhang_2023}  and a stacking analysis measurement by \citep{Stefanon_2022} at $z\simeq 8$. These previous works made use of different methods and techniques to determine the H$\alpha$ EW, such as medium/high-resolution spectroscopy, low-resolution grism spectroscopy, and narrow-band and broad-band photometry combined with SED modelling, as we did in this paper. 

Our sample of strong line emitters at $z\simeq 7 - 8$ allows us to populate a virtually unexplored part of parameter space. At those redshifts ($z\simeq 8$), only \citet{Stefanon_2022} previously obtained an estimate of the average H$\alpha$ EW$_{0}$, by median stacking 102 Lyman-break galaxies (LBG) in the 3.6$\mu$m, 4.5$\mu$m, 5.8$\mu$m, and 8.0$\mu$m bands from the \textit{Spitzer} Infrared Array Camera (IRAC). 

We also incorporate in the analysis the empirical prescriptions from \citet{Fumagalli_2012} and \citet{Faisst_2016}, who predict that the EW$_{0}$(H$\alpha$) should evolve differently below and above $z \simeq 2$. Particularly, according to the recent literature, at $z < 2$, the EW$_{0}$(H$\alpha$) should evolve as $\propto (1+z)^{1.8}$, while at $z >2$ it should evolve as $\propto (1+z)^{1.3}$. 

By inspection of  Fig. \ref{ew_z}, we can see that JWST observations at $z \geq 6$ \citep[i.e., ][and this present work]{Sun_2022, Yuanhang_2023}  suggest that the break proposed at $z\simeq 2$ in the past literature does not really hold up to such high redshifts (tiny, black, and dashed line). For that reason, we fit the evolution of EW$\mathrm{_{0}(H\alpha)}$ as a function of redshift again by considering the recent JWST observations at $z\geq 6$ as well. In this case, we find that EW$\mathrm{_{0}(H\alpha) \propto (1+z)^{2.1}}$ (bold, dark red, and dashed line in Fig \ref{ew_z}). However, larger galaxy samples are needed to confirm this finding.

Our data points are in good agreement with the stacking estimate obtained by \citet{Stefanon_2022}. Some of these values are well above the empirical median extrapolation at those redshifts, while others are consistent with it.  The prominent line emitters we analyse here constitute almost a quarter of all the MIRI-detected galaxies at $z \simeq 7-8$. The remaining MIRI sources at those redshifts should lie below the extrapolation of the empirical determination. This very large variation in the H$\alpha$ EW$_{0}$ at $z\simeq 7-8$ suggests that, even at these very high redshifts,  galaxies may be at different stages of their evolution, as we discuss in the next section.

\section{Summary and Conclusions}\label{sec_5}

In this paper, we have taken advantage of the publicly available medium and broad-band NIRCam imaging in the XDF, combined with the deepest MIRI 5.6$\mu$m imaging existing in the same field, to search for prominent (H$\beta$+[\ion{O}{3}]) and H$\alpha$ emitters at $z\simeq 7-8$. This is the first time the H$\alpha$ emission line can be detected and its flux measured in individual galaxies at such high redshifts.  This has been possible thanks to the unprecedented sensitivity of JWST observations, particularly those conducted with MIRI, for which the sensitivity gain is of more than an order of magnitude with respect to previous instruments operating at similar wavelengths \citep{Iani_2022a}. 

We found 18 galaxies which are robust candidates to be prominent (H$\beta$+[\ion{O}{3}]) emitters at  $z\simeq 7-8$, as determined from their F430M and F444W flux excess. These 18 galaxies constitute $\simeq 31\%$ of all the galaxies that we find in the XDF in the same redshift range. Among them, 16 lie on the MIRI coverage area and 12 out of 16 have a clear flux excess in the MIRI/F560W filter, indicating the simultaneous presence of a prominent H$\alpha$ emission line. The (H$\beta$+[\ion{O}{3}]) EWs$_{0}$ that we derive range from $\simeq 87.5^{+30}_{-27} \, \rm \AA$ to $2140.4^{+970}_{-154} \, \rm \AA$, with a median value of $943_{-194}^{+737} \, \rm \AA$. For  most of these galaxies, we find [\ion{O}{3}]/H$\beta > 1$, but a few have [\ion{O}{3}]/H$\beta < 1$. The two line fluxes can be separated by making use of the independent H$\alpha$ emission line measurement. This is telling us that some of the prominent (H$\beta$+[\ion{O}{3}]) emitters likely have hard radiation fields typical of low-metallicity galaxies, but not all of them.  Some are strong line emitters simply because they are intensively forming stars.

The identified H$\alpha$ emitters show an EW$_{0}$ that ranges from a few hundred to a few thousand Angstroms.  Some of these values are substantially above the expected median H$\alpha$ EW at these redshifts, as extrapolated from lower redshift determinations. We also report that, by considering the recent \textit{JWST} findings at $z\geq6$ (including this present work), $\rm EW_{0}(H\alpha)$ as a function of the redshift should evolve as follows: $\rm EW_{0}(H\alpha) \propto (1+z)^{2.1}$. However, larger samples of galaxies are needed to confirm this result. We note, however, that the prominent H$\alpha$ emitters only constitute about a quarter of all the MIRI-detected galaxies at $z\simeq 7-8$. For the remaining galaxies, the H$\alpha$ EW$_{0}$ should lie below the expected median trend. As the H$\alpha$ EW is a good proxy for the sSFR, the lower EW values could indicate that these other galaxies (the non-emitters) have either relatively low star-formation rates, or a more important underlying stellar population producing a higher continuum. This is likely the case for the non-emitters at $z\simeq 7-8$ which are relatively evolved galaxies, with best-fit ages $>10^7-10^8$ yr and stellar masses $> 10^8 \, \rm M_\odot$. 

In turn,  most of the prominent (H$\beta$+[\ion{O}{3}]) and H$\alpha$ emitters are characterised by higher sSFR, with basically all of them being starburst galaxies or on the way to/from the starburst cloud. The majority of the prominent (H$\beta$+[\ion{O}{3}]) emitters are very young galaxies (best-fit ages $<10^7 \, \rm yr$), so they might be in their first major star-formation episode. A few others are almost as old as the Universe at their redshifts and have already built significant stellar mass ($> 10^8 \, \rm M_\odot$), suggesting that they may be experiencing a rejuvenation effect. 

Therefore, the overall conclusion of this work is that the galaxies present at the EoR are likely at different stages of their evolution. And strong line emission is present in a minor, but significant fraction of sources. 

Considering the H$\alpha$ fluxes inferred for the prominent H$\alpha$ emitters, we estimated their contribution to the cosmic SFRD at $z\simeq 7-8$. We found $\mathrm{log_{10} (\rho_{SFR_{H\alpha}}/(M_{\odot}\;yr^{-1}\;Mpc^{-3})) \simeq -2.35 \pm 0.3}$, in excellent agreement with independent measurements from the literature based on rest-frame UV luminosities, and with theoretical predictions and empirical extrapolations from lower redshifts. We note, however, that this estimated SFRD must be considered a lower limit, as it only takes into account the most prominent  H$\alpha$ emitters at $z\simeq 7-8$.
We also considered the SFR$\mathrm{_{UV}}$ for all the other galaxies at $z\simeq 7-8$  to obtain a total SFRD value at that redshift interval. We concluded that the strong H$\alpha$ emitters produced about a quarter of the total SFRD at $z\simeq 7-8$, which suggests that they likely have had a significant role in the process of reionization. In a future paper, we will conduct a more detailed investigation of these sources, in order to better understand their nature.

\acknowledgments
In memoriam to the MIRI European Consortium members Hans-Ulrik Noorgard-Nielsen and Olivier Le F\`evre.

The authors would like to acknowledge an anonymous referee for a careful reading and useful comments on this manuscript.
This work is based on observations made with the NASA/ESA/CSA James Webb Space Telescope. The data were obtained from the Mikulski Archive for Space Telescopes at the Space Telescope Science Institute, which is operated by the Association of Universities for Research in Astronomy, Inc., under NASA contract NAS 5-03127 for JWST. These observations are associated with programs GO \#1963, GO \#1895 and GTO \#1283. The authors acknowledge the team led by coPIs C. Williams, M. Maseda and S. Tacchella, and PI P. Oesch, for developing their respective observing programs with a zero-exclusive-access period. Also based on observations made with the NASA/ESA Hubble Space Telescope obtained from the Space Telescope Science Institute, which is operated by the Association of Universities for Research in Astronomy, Inc., under NASA contract NAS 5–26555. The specific observations analyzed can be accessed via:\dataset[DOI: 10.17909/T91019]..
The work presented here is the effort of the entire MIRI team and the enthusiasm within the MIRI partnership is a significant factor in its success. MIRI draws on the scientific and technical expertise
of the following organisations: Ames Research Center, USA; Airbus Defence and Space, UK; CEA-Irfu, Saclay, France; Centre Spatial de Li\`{e}ge, Belgium;
Consejo Superior de Investigaciones Cient\'{\i}ficas, Spain; Carl Zeiss Optronics, Germany; Chalmers University of Technology, Sweden; Danish Space Research
Institute, Denmark; Dublin Institute for Advanced Studies, Ireland; European Space Agency, Netherlands; ETCA, Belgium; ETH Zurich, Switzerland; Goddard Space Flight Center, USA; Institute d’Astrophysique Spatiale, France; Instituto Nacional de T\'ecnica Aeroespacial, Spain; Institute for Astronomy, Edinburgh, UK; Jet Propulsion Laboratory, USA; Laboratoire d’Astrophysique de Marseille (LAM), France; Leiden University, Netherlands; Lockheed Advanced
Technology Center (USA); NOVA Opt-IR group at Dwingeloo, Netherlands; Northrop Grumman, USA; Max-Planck Institut für Astronomie (MPIA), Heidelberg, Germany; Laboratoire d’Etudes Spatiales et d’Instrumentation en Astrophysique (LESIA), France; Paul Scherrer Institut, Switzerland; Raytheon Vision Systems, USA; RUAG Aerospace, Switzerland; Rutherford Appleton Laboratory (RAL Space), UK; Space Telescope Science Institute, USA; Toegepast-
Natuurwetenschappelijk Onderzoek (TNO-TPD), Netherlands; UK Astronomy Technology Centre, UK; University College London, UK; University of Amsterdam, Netherlands; University of Arizona, USA; University of Cardiff, UK; University of Cologne, Germany; University of Ghent; University of Groningen, Netherlands; University of Leicester, UK; University of Leuven, Belgium; University of Stockholm, Sweden; Utah State University, USA.

KIC and EI acknowledge funding from the Netherlands Research School for Astronomy (NOVA). KIC, RNC and VK acknowledge funding from the Dutch Research Council (NWO) through the award of the Vici Grant VI.C.212.036. The Cosmic Dawn Center is funded by the Danish National Research Foundation under grant No. 140. LC acknowledges financial support from Comunidad de Madrid under Atracci\'on de Talento grant 2018-T2/TIC-11612. SG acknowledges the support of the Cosmic Dawn Center of Excellence funded by the Danish National Research Foundation under grant 140. G.\"O., A.B. \&  J.M.  acknowledge support from the Swedish National Space Administration (SNSA). AAH acknowledges support from PID2021-124665NB-I00 funded by the Spanish Ministry of Science and Innovation and the State Agency of Research MCIN/AEI/10.13039/501100011033. J.H. and D.L. were supported by a VILLUM FONDEN Investigator grant to J.H. (project number 16599).

JAM and ACG acknowledge support by grant PIB2021-127718NB-100 by the Spanish Ministry of Science and Innovation/State Agency of Research MCIN/AEI/10.13039/ 501100011033 and by “ERDF A way of making Europe”.

PGP-G acknowledges support from Spanish Ministerio de Ciencia e Innovaci\'on MCIN/AEI/10.13039/501100011033 through grant PGC2018-093499-B-I00.

JPP and TVT acknowledge funding from the UK Science and Technology Facilities Council, and the UK Space Agency.

\vspace{5mm}
\facilities{{\sl HST}, {\sl JWST}}.

\software{\texttt{Astropy} \citep{astropy_2018}, 
          \texttt{LePHARE} \citep{LePhare_2011},
          \texttt{NumPy} \citep{Numpy},
          \texttt{pandas} \citep{Pandas}
          \texttt{Photutils} \citep{Photutils}, 
          \texttt{SciPy} \citep{Scipy}
          \texttt{Source Extractor} \citep{SExtractor},
          \texttt{TOPCAT} \citep{Topcat}.
          }

\bibliography{References}{}
\bibliographystyle{aasjournal}
\end{document}